\documentclass[a4paper,11pt]{article}
\pdfoutput=1
\usepackage{jheppub}

\usepackage{graphicx}
\usepackage{placeins}
\usepackage{url}
\usepackage{multirow}
\usepackage{amssymb}
\usepackage{amsmath}
\usepackage{mathtools}
\usepackage{makecell}
\usepackage{multirow}
\usepackage{bbm}
\usepackage{xspace}
\usepackage{booktabs}
\usepackage{xcolor}
\usepackage{cleveref}
\usepackage[symbol]{footmisc}
\usepackage{seqsplit}
\usepackage[htt]{hyphenat}
\usepackage{microtype}

\newcommand{\val}[2]{$#1 \,{\scriptstyle\pm}\, #2$}
\newcommand{\bestval}[2]{\boldmath$#1 \,{\scriptstyle\pm}\, #2$}

\usepackage{algorithm}
\usepackage[noend]{algpseudocode}

\begin{document}

\title{Explicit or Implicit? \\ Encoding Physics at the Precision Frontier}

\affiliation[a]{Department of Physics, Harvard University, Cambridge, MA 02138, USA}
\affiliation[b]{Department of Physics and Astronomy, University of California, Irvine, California, USA}
\affiliation[c]{Nagoya University, Kobayashi-Maskawa Institute, Aichi 464-8602, Japan}
\affiliation[d]{Department of Particle Physics and Astrophysics, Stanford University, Stanford, CA 94305, USA}
\affiliation[e]{Fundamental Physics Directorate, SLAC National Accelerator Laboratory, Menlo Park, CA 94025, USA}
\affiliation[f]{Institute for Theoretical Physics, Universität Heidelberg, Germany}
\affiliation[g]{Interdisciplinary Center for Scientific Computing (IWR), Universität Heidelberg, Germany}

\author[a]{Victor Bresó-Pla,}
\author[b]{Kevin Greif,}
\author[c]{Vinicius Mikuni,}
\author[d,e]{Benjamin Nachman,}
\author[f,g]{Tilman Plehn,}
\author[d]{Tanvi Wamorkar,}
\author[b]{Daniel Whiteson}

\abstract{
High-performance machine learning tools in particle physics rest on two complementary directions: encoding symmetries explicitly in the architecture, and implicitly learning the structure of the data through large-scale (pre-) training.
We compare the performance of the representative L-GATr and \textsc{OmniLearn} models on three especially challenging tasks: reweighting-based unfolding, likelihood-ratio estimation, and weakly supervised anomaly detection.
Across all benchmarks, both methods achieve comparable performance given the statistical precision of the finetuning datasets, suggesting that the significant efficiency gains from encoding known particle physics structures are largely method-independent.
}

\maketitle

\clearpage
%%%%%%%%%%%%%%%%%%%%%%%%%%%%%%%%%%%%%%%%%%%%%%%%%%%%%%%%%%%%

\section{Introduction}
\label{intro}
Modern machine learning (ML) is becoming an essential tool in particle physics, transforming tasks such as data acquisition, object reconstruction, event characterization, process and detector simulation, cross section measurements, parameter estimation, and anomaly detection~\cite{Radovic:2018dip,Karagiorgi:2021ngt,Feickert:2021ajf,Plehn:2022ftl}. 
%ranging from event classification and unfolding to anomaly detection and new physics searches. 
A central aspect of ML in particle physics is encoding
%representation learning, the question of how to improve the training efficiency and the network performance by using 
the known structures governing the data, such as symmetries, in a way that improves performance and robustness while also being data efficient.
%the many structures and symmetries governing the data.
%, which is different than natural language or images.  
%For example, the variable multiplicity in position or momentum space naturally lends itself to a point-cloud representation. 
Quantum field theory dictates the unique symmetries underlying particle physics data, which are often naturally represented as point clouds.  How to best integrate such structures is an active area of research~\cite{Bogatskiy:2022hub}.

Two general strategies for incorporating physics knowledge into ML models have emerged. One approach is to explicitly incorporate physics structures into the model setup, either explicitly through the network architecture or as part of the training~\cite{AkhoundSadegh:2023xlp,Elhag:2024oer,Nabat:2024nce,Hebbar:2025adf}. An example is Lorentz-equivariant networks~\cite{Bogatskiy:2020tje,Villar:2021hph,Gong:2022lye,Qiu:2022xvr,Bogatskiy:2022hub,Bogatskiy:2022hrp,Li:2022xfc,Hao:2022zns,Bogatskiy:2023nnw,Batatia:2023nqg,Brehmer:2024yqw,Favaro:2025pgz,Bahl:2024gyt}. The motivation of explicit equivariance is that models do not need to learn known critical structures of the input data, the cost is that the models themselves tend to be more complex~\cite{Bogatskiy:2023fug,Petitjean:2025zjf}.

An alternative approach is to incorporate physical reasoning implicitly 
through large-scale pretraining on diverse datasets that represent similar structures. Here, models learn appropriate representations through pretraining, and these representations are subsequently adapted to downstream tasks through finetuning~\cite{Kuchera:2018msh,Dillon:2021gag,Dillon:2022tmm,Chappell:2022yxd,Dreyer:2022yom,Beauchesne:2023xhj,Dillon:2023zac,Birk:2024knn,Harris:2024sra,Golling:2024abg,Leigh:2024ked,Hallin:2025wme,Bardhan:2025dpn}. The implicit approach shifts the focus from modifying the architecture to curation of large-scale training datasets, whereas the architectures mostly require substantial capacity to learn the physical attributes of the data.

We provide a realistic comparison of explicit and implicit approaches using two state-of-the-art models --- the Lorentz Geometric Algebra Transformer (L-GATr)~\cite{brehmer2023geometricalgebratransformer,Brehmer:2024yqw,Petitjean:2025zjf} and \textsc{OmniLearn}~\cite{Mikuni:2024qsr,Mikuni:2025wjk,Bhimji:2025tpq,Mikuni:2025ezi,Elsharkawy:2026bnm}, a foundation model that acquires physics priors implicitly through large-scale pretraining. Previous studies have compared these approaches for top tagging~\cite{Kasieczka:2019dbj} or multi-label jet classification~\cite{Qu:2022mxj}, where the classes are quite different.  %The goal in this case is to achieve high accuracy so that data can be enriched in the target class after a suitable event selection.  
These studies indicate that equivariance yields substantial improvements in accuracy, data efficiency, and robustness compared to non-equivariant models with comparable parameter counts, but the best performance is reached by pretrained transformers~\cite{Brehmer:2024yqw,Bhimji:2025tpq}.

In the precision collider physics program, there is a growing need for ML models that can faithfully capture very subtle differences between two classes.  
%Unlike previous studies\footnote{In both cases, the likelihood ratio is the solution, but its form qualitatively differs between task types.}, the goal here is not necessarily high classification accuracy.  
Tasks such as simulation-based inference~\cite{doi:10.1073/pnas.1912789117}, unfolding~\cite{Canelli:2025ybb}, and anomaly detection~\cite{Kasieczka:2021xcg,Aarrestad:2021oeb,Belis:2023mqs} seek high-fidelity likelihood ratio estimation. The goal of this paper is to compare explicit and implicit approaches in this context of nearly identical classes.  

The paper is organized as follows. Section~\ref{priors} describes our two representative models.
%we study for explicit and implicit physics priors.
%with Section~\ref{lgatr} describing the Lorentz Geometric Algebra Transformer architectures and Section~\ref{omnilearn} describing the Omnilearn architecture. 
%Section~\ref{compute} compares the two approaches in terms of their computational resource management. 
Section~\ref{comparison} then analyzes their performance in classification tasks involving similar classes, namely reweighting-based unfolding for $pp$ collisions in Section~\ref{zjets}, likelihood ratio estimation for $ep$ collisions in Section~\ref{h1}, and weakly-supervised anomaly detection in Section~\ref{ad}. We close with an outlook in Section~\ref{conclusion}.

%%%%%%%%%%%%%%%%%%%%%%%%%%%%%%%%%%%%%%%%%%%%%%%%%%%%%%%%%%%%
\section{Explicit versus implicit physics knowledge}
\label{priors}
Collider physics data are defined in a highly structured space that respects a number of known exact or approximate symmetries. This allows us to compare two ways of including them in neural networks, either explicitly through constrained data representation and operation space, or implicitly as universally learned data features. 
%
%Crucially, both approaches can learn the breaking patterns of approximate symmetries.
%, with explicit methods requiring deliberate architectural design choices and implicit methods learning them directly from the data. 
%Furthermore, both strategies can learn additional structures that cannot be encoded in a symmetry group. 
Methods with explicit priors factorize out the known symmetry structure and methods with implicit priors achieve this through large-scale data exposure.  As a result, both approaches can learn the breaking patterns of approximate symmetries and can learn additional structures that cannot be encoded in a symmetry group. 
We use the Lorentz Geometric Algebra Transformer (L-GATr) and \textsc{OmniLearn} as representatives and cutting-edge examples of the explicit and implicit methods, respectively.

%%%%%%%%%%%%%%%%%%%%%%%%%%%%%%%%%%%%%%%%%%%%%%%%%%%%%%%%%%%%
\subsection{Lorentz-equivariant transformer}
\label{lgatr}
\label{lgatr}
L-GATr~\cite{brehmer2023geometricalgebratransformer,Brehmer:2024yqw,Petitjean:2025zjf} is a transformer architecture equivariant under Lorentz transformations $\Lambda$ of phase space points $x$,
\begin{align}
    \text{L-GATr}(\Lambda(x))=\Lambda(\text{L-GATr}(x)) \; .
\end{align}
It embeds its inputs into the space-time geometric algebra. In this representation, their geometric structure is manifest, and operations on them can be restricted to operations that preserve Lorentz symmetry, with the option of learning an explicit breaking. The elements of the algebra are called multivectors, which can be written as
\begin{align}
    x = x^S \; 1 
    + x^V_\mu  \; \gamma^\mu 
    + x^B_{\mu\nu} \; \sigma^{\mu\nu} 
    + x^A_\mu \; \gamma^\mu \gamma^5 
    + x^P \; \gamma^5 
    \qquad \text{with} \qquad 
    \begin{pmatrix}
    x^S \\ x^V_\mu \\ x^B_{\mu\nu} \\ x^A_\mu \\ x^P 
    \end{pmatrix} 
    \in\mathbb{R}^{16} \; .
\label{eq:multivector}
\end{align}
This expression combines scalars $x^S$, vectors $x^V_{\mu}$, bi-vectors, $x^B_{\mu\nu}$, axial vectors $x^A_{\mu}$, and pseudoscalars $x^P$. 

The vector space is expanded to the geometric algebra through a geometric product, in analogy to an outer product. It constructs higher-order geometric objects through an antisymmetric product of the algebra elements. Specifically, vectors can be paired antisymmetrically to define the new basis elements $\sigma^{\mu\nu}$, $\gamma^{\mu}\gamma^5$ and $\gamma^5$. Together with the scalar basis element $1$, they define a decomposition of multivectors in grades, encoding the dimension of the objects. The notation reflects the close relation of the spacetime geometric algebra with the Dirac algebra, where both share the operational and representation properties, but the Dirac algebra is defined over complex space.

The multivector representation provides a simple way to enforce equivariance for a wide range of network operations and across many network architectures. The only constraint that needs to be imposed is that a given network acts independently on each multivector grade, as every grade transforms under a different representation of the Lorentz group. L-GATr applies this representation to transformer architectures, so it can take advantage of the desirable scaling and expressivity of transformers.    

In practice, L-GATr organizes the data as a collection of tokens and encodes their geometric content into a multivector. For jet physics, the tokens correspond to the constituents, each in a multivector representation following from the 4-vector input,
\begin{align}
    x^V_{\mu} = p_{\mu}
    \qquad \text{and} \qquad 
    x^S = x^T_{\mu\nu}=x^A_\mu=x^P=0 \; . 
    \label{eq:embedding}
\end{align}
%
%Thus, constituent features in a $(E, p_x, p_y, p_z)$ format must always be provided as part of the input. 
L-GATr can also process constituent features not contained in the 4-vector and separately from the multivector representation, with interactions in the linear and attention layers. This allows the network to use features such as the particle-ID, the transverse momentum $p_\textrm{T}$ or the rapidity $y$,  at the cost of a potential equivariance breaking. Given this input structure, we can limit ourselves to Lorentz scalars and vectors for many collider physics applications using the L-GATr-slim implementation~\cite{Petitjean:2025zjf}. 

Finally, L-GATr can accommodate any additional information not linked to a specific constituent through extra tokens. We will use such tokens to process event-level features and use a one-hot encoding to distinguish them from the constituent-level features. 
% Both constituent-level and jet-level scalar features are included for each item in the sequence, but we mask to zero the jet-level inputs for the constituents and the constituent-level inputs for the jets.
The extra tokens can circumvent  Lorentz equivariance, an essential feature whenever Lorentz symmetry is only partially preserved. Symmetry-breaking tokens are engineered so that any operation involving them breaks the equivariance down to the subgroup that leaves the new input multivectors invariant. This mechanism also covers scalar inputs, where the extra features will induce symmetry breaking down to subgroups for which the input selection behaves like Lorentz scalars. This enables tunable and dynamic symmetry breaking without adjusting the architecture~\cite{Favaro:2025pgz}. 

The output of the L-GATr transformer requires an extra step that undoes multivector embedding. Throughout this study, the L-GATr output should be a scalar for each jet or event, so we add an extra class token and select its scalar component as the network output.

With this architecture, L-GATr achieves excellent performance in several tasks~\cite{Brehmer:2024yqw},  such as amplitude regression, jet tagging, event generation, and generative unfolding~\cite{Petitjean:2025tgk}. The equivariance allows us to work with relatively small datasets to reach competitive performance. The cost is an increased memory use due to the size of the multivector representation.
%, which also impacts the size of the hidden states and outputs with respect to a non-equivariant network. 
It can be reduced by restricting the algebra grades to scalars and vectors and removing the geometric product~\cite{Petitjean:2025zjf}. We test this L-GATr-slim version together with the full model in Section~\ref{zjets} and analyze the computational costs of L-GATr, L-GATr-slim, and \textsc{OmniLearn} for jet classification in Appendix~\ref{compute}. Instead of L-GATr, we could also use the LLoCa realization of a Lorentz-equivariant transformer~\cite{Favaro:2025pgz}, but we do not expect a qualitatively different result.

%%%%%%%%%%%%%%%%%%%%%%%%%%%%%%%%%%%%%%%%%%%%%%%%%%%%%%%%%%%%
\subsection{OmniLearn}
\label{omnilearn}
% \label{omnilearn}
\textsc{OmniLearn}~\cite{Mikuni:2024qsr} is a foundation model whose backbone is based on the Point-Edge Transformer (PET), a hybrid architecture combining transformer blocks with graph network layers. PET uses a shared representation learned from large-scale pretraining, whose output is subsequently passed to task-specific heads. 
%The \textsc{OmniLearn} model is described in detail in Ref.; here we summarize its key characteristics.
The model is pretrained on approximately $10^8$ jets from the JetClass dataset~\cite{Qu:2022mxj}. Through this large-scale pretraining, \textsc{OmniLearn} learns general representations of jets that capture correlations among constituents and characteristic substructure patterns.

The PET backbone processes constituent 4-vectors for each jet, treating jets as unordered sets of constituents. The constituent features are conditioned on diffusion time parameters and augmented with local geometric information via graph convolution layers. To accommodate heterogeneous input features, always including kinematic information but not always features such as particle ID or vertex information, the model employs a feature dropout. In this approach, additional input features are randomly dropped with fixed probabilities and replaced with zeros during training, enabling the model to learn robust representations in the presence and absence of auxiliary information.

The output of the PET backbone is provided to a classifier head and a generative head. The classifier head operates on the constituent-level representations and incorporates additional global jet kinematic information. The generative head, used for diffusion-based generation, similarly consumes constituent and jet information, while explicitly incorporating diffusion time conditioning. This dual-head design enables \textsc{OmniLearn} to be jointly trained on classification and generative tasks. 

% \textsc{OmniLearn} represents an implicit approach to incorporating physics knowledge through large-scale pretraining, in contrast to models such as L-GATr that encode physical symmetries explicitly at the architectural level. Rather than enforcing physics constraints by construction, \textsc{OmniLearn} develops physics-informed internal representations through exposure to a diverse and extensive training dataset. These representations can then be fine-tuned for new tasks with minimal adaptation, typically requiring modification only of the final task-specific layers.

The foundation-model approach offers significant flexibility: no architectural redesign is required when adapting the model to different collision systems or detector configurations, as the model adapts through fine-tuning. The effectiveness of this approach has been demonstrated in systematic transfer learning studies, where \textsc{OmniLearn} achieves state-of-the-art performance on binary classification such as top tagging and quark–gluon discrimination, while requiring 50–70\% fewer training epochs compared to models trained from scratch.

\textsc{OmniLearn} has a substantial one-time computational cost, pretraining with 128 GPUs for 20 epochs on $10^8$ jets. However, this leads to a significant efficiency gain in terms of reduced training time for downstream tasks. Fine-tuned models typically converge 2 to 3.5 times faster than comparable architectures without pretraining. We quantify the computational cost 
%of pretraining and fine-tuning and compare it with L-GATr 
in Appendix~\ref{compute}.  The performance may be further improved with bigger models and more pre-training data~\cite{Bhimji:2025tpq}.

\clearpage
%%%%%%%%%%%%%%%%%%%%%%%%%%%%%%%%%%%%%%%%%%%%%%%%%%%%%%%%%%%%
\section{Precision classification with similar classes}
\label{comparison}
\label{classification}
Although networks in collider physics have been shown to improve with an appropriately scaled increase in network size, training dataset size, training time, GPU investment, and energy consumption~\cite{Batson:2023ohn,Vigl:2026ppx}, this scaling also leads to significant cost on the application side, in terms of memory and evaluation time. This is why our study instead focuses on efficiency gains from explicitly or implicitly using physics information, at comparable and realistic training and application cost. 

Our precision classification tasks are based on likelihood ratio estimations, but differ in the relationship between classes, the physical processes involved, and the downstream applications. Sections~\ref{zjets} and~\ref{h1} use the classifier output to reweight simulations to data distributions for LHC and HERA observables, using the \textsc{OmniFold} unfolding method~\cite{Andreassen:2019cjw,Andreassen:2021zzk}. Section~\ref{ad}  focuses on distinguishing two data distributions that differ only by the presence of a small fraction of anomalous events in one of them. This task is relevant for new physics searches where the effect of the new physics phenomena on the data is expected to be very small.   

%%%%%%%%%%%%%%%%%%%%%%%%%%%%%%%%%%%%%%%%%%%%%%%%%%%%%%%%%%%%
\subsection{Reweighting-based unfolding for $pp$ collisions}
\label{zjets}
\label{zplusjets}
Unfolding statistically corrects for unwanted distortions in the data. This task requires an accurate forward model of the effects to be corrected.  Once unfolded, the data can be readily compared between experiments and with theoretical predictions.
%Unfolding is an application of representation learning, where recorded data is reported in a pre-defined representation, typically defined through a numerical forward simulation. It allows experiments to choose an optimal and efficient data representation for a given analysis and even publish data for analyses outside a collaboration. 
%One way to represent unfolded data is to statistically adjust measured spectra to account for distortions introduced, for instance, by the detector. 
In our application, we unfold detector effects and represent data in terms of particles as objects entering the detector.
This is a common step in many measurements of Standard Model processes in collider physics.
Unfolding requires the observed data and a matched sample of simulated events before and after the corresponding forward simulation.
For our example of detector unfolding, these configurations are referred to as particle level and detector level.

Traditional unfolding methods bin the data and simulation and adjust the resulting histograms~\cite{Hocker:1995kb,DAGOSTINI1995487,Cowan:2002in,Blobel:2011fih,Schmitt:2012kp,Behnke:2013pga,Brenner:2019lmf}.
ML-unfolding allows for the correction of high-dimensional distributions without binning~\cite{Andreassen:2019cjw,Bellagente:2019uyp,Bellagente:2020piv,Andreassen:2021zzk,Huetsch:2024quz}. For \textsc{OmniFold}~\cite{Andreassen:2019cjw,Andreassen:2021zzk} we train a classifier to re-weight a sample of simulated events to represent the unfolded data. This is done through an iterative procedure that requires training two classifiers per iteration, although a more direct approach might be feasible~\cite{Ore:2026qgp}. By assumption, the data and simulation are very similar, providing us with a perfect example for classification with nearly identical classes.

% \textsc{\textsc{OmniFold}} works iteratively, at each step a classifier is trained from scratch on a dataset where the weights from the prior iteration are used to reweight the simulation events closer to the data distribution.

We compare L-GATr and \textsc{OmniLearn} as re-weighting classifiers on a  benchmark dataset containing simulated collisions for
\begin{align}
 pp \to Z + \text{jets} 
 \qquad \text{with} \qquad 
 p_{\textrm{T},Z} > 200 \text{GeV} \, 
\end{align} 
at 14~TeV~\cite{OmniFold_zenodo,Andreassen:2019cjw}. The dataset consists of two sets of events generated by different simulators, each at detector and particle levels. The first set is generated with Herwig 7.1.5~\cite{Bahr:2008pv,Bellm:2015jjp,Bellm:2017bvx} with the default tune and is used as `data'. The second set is generated with \textsc{Pythia8}~\cite{Sjostrand:2014zea} with Tune 26~\cite{pythiatune} and is used as `simulation'. Both datasets use Delphes~\cite{deFavereau:2013fsa,Selvaggi:2014mya,Mertens:2015kba} and the particle flow reconstruction CMS tune to simulate detector effects. The jet content consists of particle flow objects at detector level and stable truth particles at particle level. Jets are clustered with the anti-$k_T$ algorithm~\cite{Cacciari:2005hq,Cacciari:2008gp,Cacciari:2011ma} with $R=0.4$, as implemented in FastJet 3.3.2~\cite{Cacciari:2011ma}. 
% The jet momenta in this dataset range between 500-1000 GeV, which is much lower than the average momentum in the JetClass dataset.

% We perform the comparison for the tasks of re-weighting the detector level simulation to match data (step one of the first iteration of \textsc{OmniFold}) and re-weighting the particle level simulation to match the truth data distribution.
% We study two tasks relevant for \textsc{OmniFold}: the reweighting of the 'simulation' distribution into the 'data' distribution at reconstructed level and the complete unfolding after $n=5$ iterations.

%-------------------------------------------------
\begin{table}[t]
%\hspace{-0.8cm}
\small
\centering
\begin{tabular}{l c c}
\toprule
Input & Scalar & 4-vector \\
\midrule
Constituents &
\makecell{
$\Delta \eta_{i,\text{jet}}$, $\Delta \phi_{i,\text{jet}}$, $\log p_\textrm{T}$, $\log E$, \\
$\log \!\left(1-\frac{p_{\textrm{T},i}}{p_{\textrm{T},\text{jet}}}\right)$,
$\log \!\left(1-\frac{E_i}{E_{\text{jet}}}\right)$, 
$\Delta R_{i,\text{jet}}$, charge
}
& $E$, $p_x$, $p_y$, $p_z$ \\[7mm]
Jets &
$p_\textrm{T}$, $\eta$, $m$, multiplicity & $E$, $p_x$, $p_y$, $p_z$ 
\\
\bottomrule
\end{tabular}
\caption{Input features for L-GATr and L-GATr-slim on the $Z$+jets dataset and for L-GATr on the H1 dataset. For the $Z$+jets dataset, $f(\text{PID})$ is additionally included as a scalar feature for jet constituents. The variables listed in the scalar column are the input features used to train \textsc{OmniLearn} on both tasks.}
\label{tab:zjets_inputs}
\end{table}
%-------------------------------------------------

\textsc{\textsc{OmniLearn}} uses constituent-level and jet-level inputs. The input for the classifier training is summarized in the scalar column in Table~\ref{tab:zjets_inputs}. Its training follows the exact setup described in Ref.~\cite{Mikuni:2024qsr} and uses the pretrained network weights provided in the public data release as the starting point for the fine-tuning. We could have used previous \textsc{OmniLearn} results, but the change in the \textsc{Pythia8} tune used in this study required retraining it for a trustworthy comparison. We also train the PET network from scratch using the same configuration, to quantify the impact of pretraining. 
% In the case of L-GATr, the training on both tasks starts from scratch and matches the \textsc{\textsc{OmniLearn}} setup as closely as possible. This entails matching the data inputs, training iterations and network size, which requires careful consideration due to the inherent design differences between the two architectures. 

L-GATr is trained from random weight initializations and matches \textsc{OmniLearn} in inputs, training epochs, and network size. The scalar inputs in Table~\ref{tab:zjets_inputs} are standardized and given to the scalar channels, and the constituent and jet 4-momenta are added as multivector inputs.  We also test a minimal setup, where we train both networks using only constituent vectors. The number of parameters in L-GATr is either $10^6$ or $2 \times 10^6$, where the former is roughly the size of the PET body used in \textsc{OmniLearn} without task-specific heads, and the latter is the total \textsc{OmniLearn} size. L-GATr-slim only uses $7\times 10^5$ parameters. We observe that all tested L-GATr versions reach their maximum performance with at most 10 training epochs, which matches the amount of fine-tuning steps that \textsc{OmniLearn} takes to reach convergence on all studied tasks~\cite{Mikuni:2024qsr}. For a simple hyper-parameter optimization of the L-GATr networks, we start with the \textsc{OmniLearn} hyper-parameters and then optimize the batch size, learning rate, number of attention blocks, and weight decay parameter, as described in  Appendix~\ref{hyperparameters_zjets}.
% As for the hyperparameters, we use the setup used in \textsc{\textsc{OmniLearn}} as a starting reference and from there we perform a sweep over batch size, learning rate scheduler, network shape and weight decay.
% We also study the importance of including the scalar inputs for L-GATr by training a set of networks only on the 4-momenta features. 

First, we compare the performance of L-GATr, L-GATr-slim, the PET network trained from scratch, and \textsc{OmniLearn} on the first step of the first iteration of \textsc{OmniFold}, which we call reweighting. The unfolding task is to reweight the `simulated' events to the `data' distribution at detector level. We observe a large variability in the performance of L-GATr on repeated trainings across a wide range of hyper-parameter setups, which complicates network optimization. L-GATr-slim is also unstable under multiple runs, but its behavior is much more predictable under hyper-parameter changes.   We show the best-performing trainings from L-GATr and L-GATr-slim and give the full details about our training setup in Appendix~\ref{gatr_details}. The results from \textsc{OmniLearn} are stable against repeated trainings. 

%-------------------------------------------------
\begin{figure}[t]
    \hspace{-26pt}
    \setlength{\tabcolsep}{-5pt}
    \renewcommand{\arraystretch}{0}
    \begin{tabular}{@{}ccc@{}}
    \includegraphics[width=0.39\textwidth]{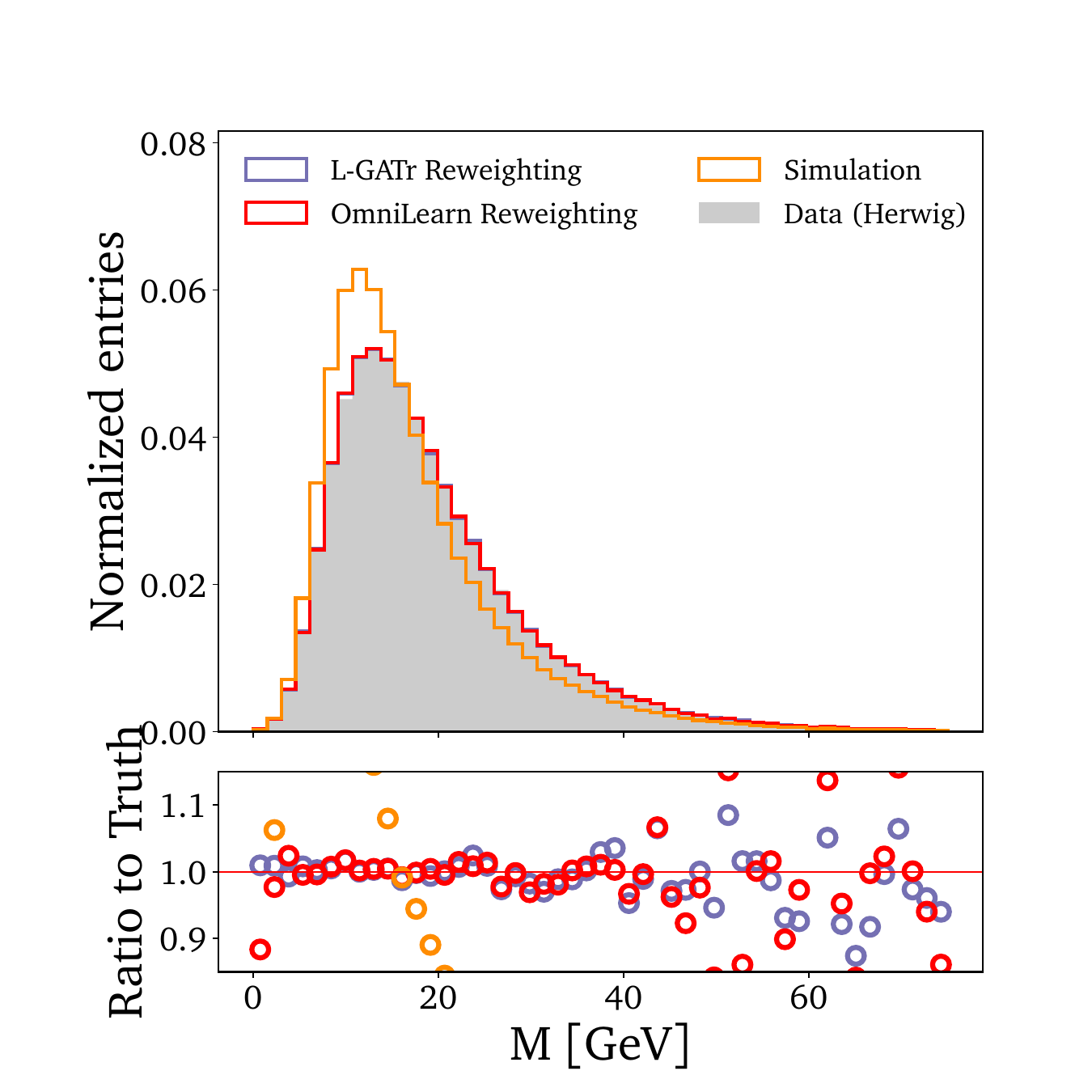}% &
    \hspace{-16pt}%
    \includegraphics[width=0.39\textwidth]{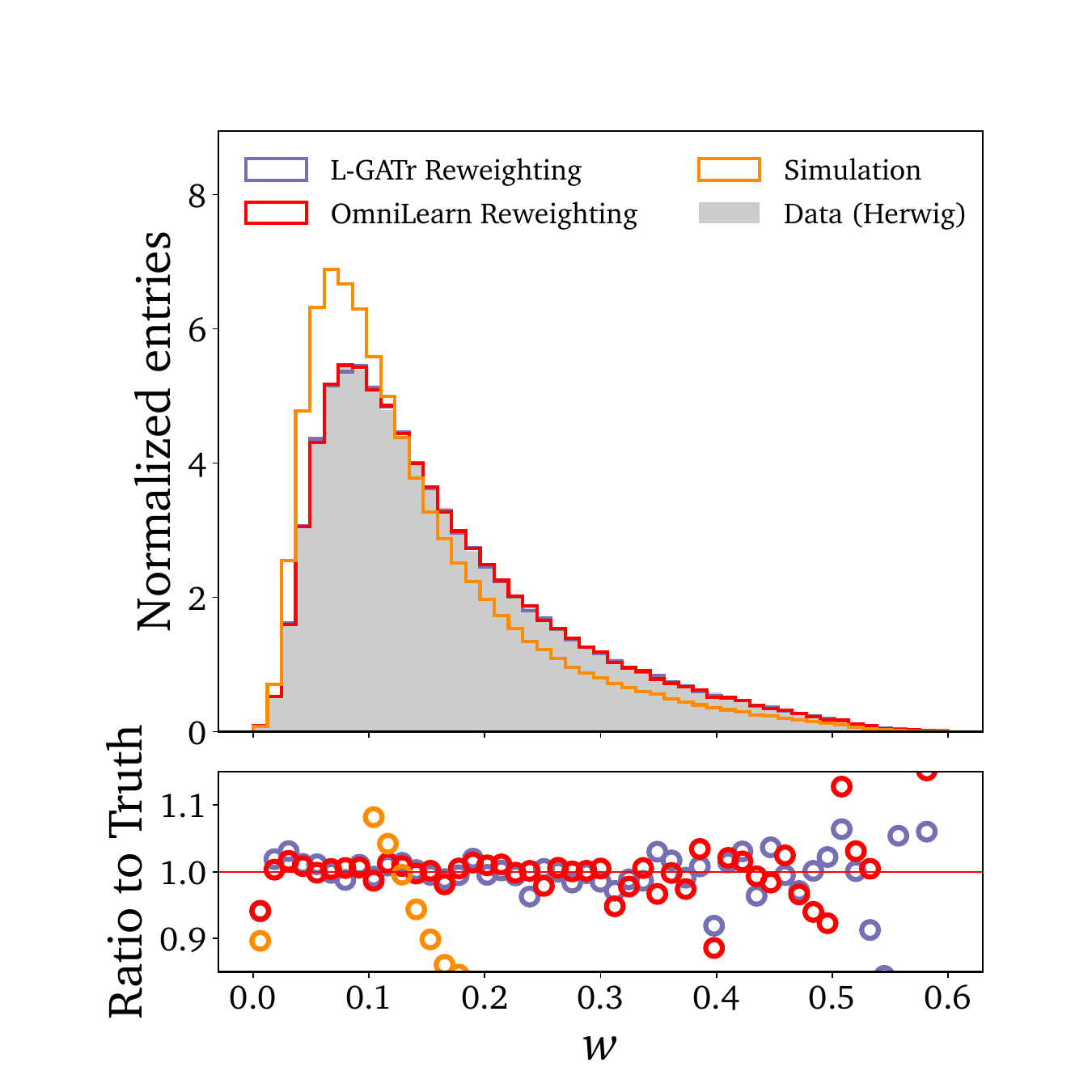}% &
    \hspace{-16pt}%
    \includegraphics[width=0.39\textwidth]{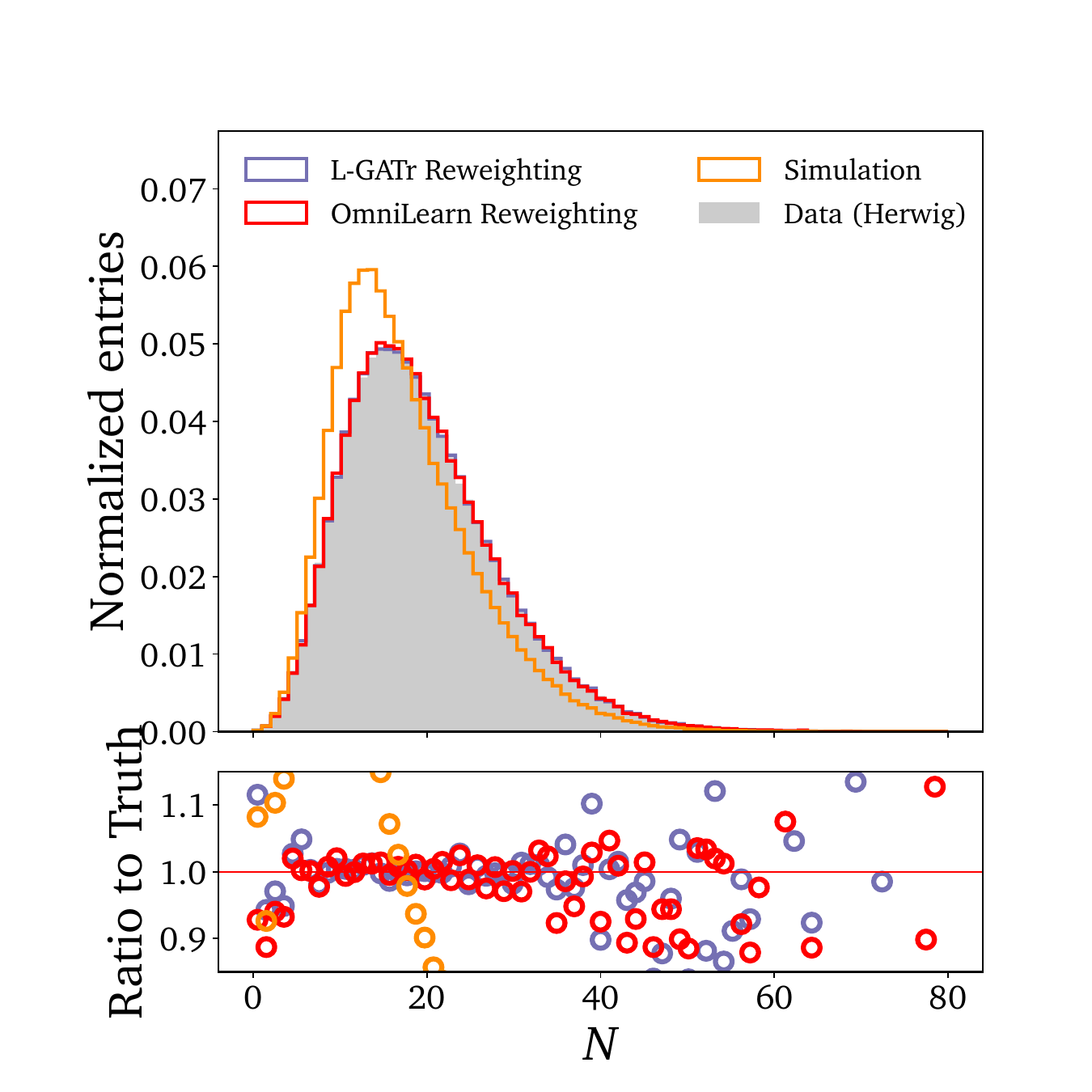} \\%[-2mm]
    \includegraphics[width=0.39\textwidth]{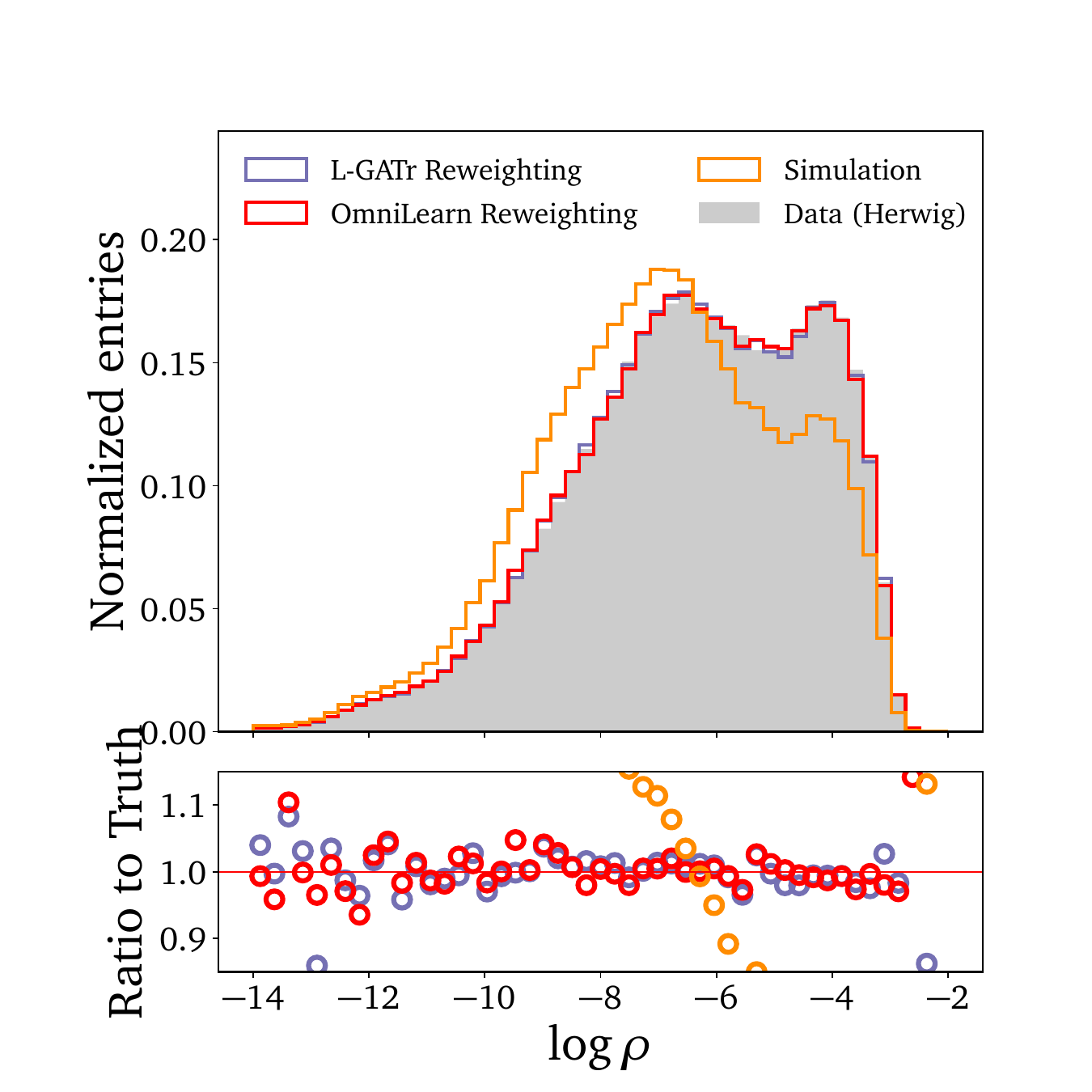}% &
    \hspace{-16pt}% 
    \includegraphics[width=0.39\textwidth]{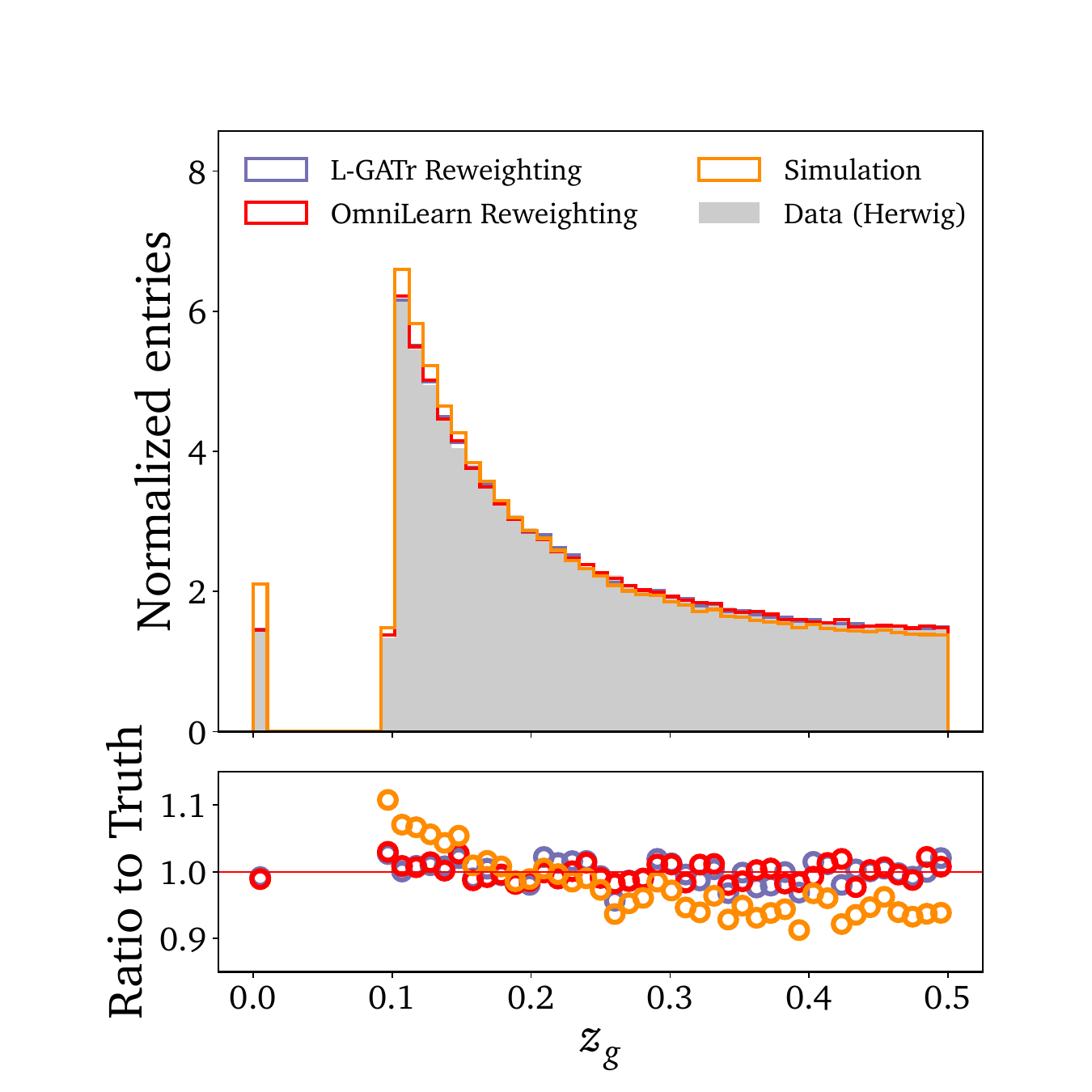}% &
    \hspace{-16pt}%
    \includegraphics[width=0.38\textwidth]{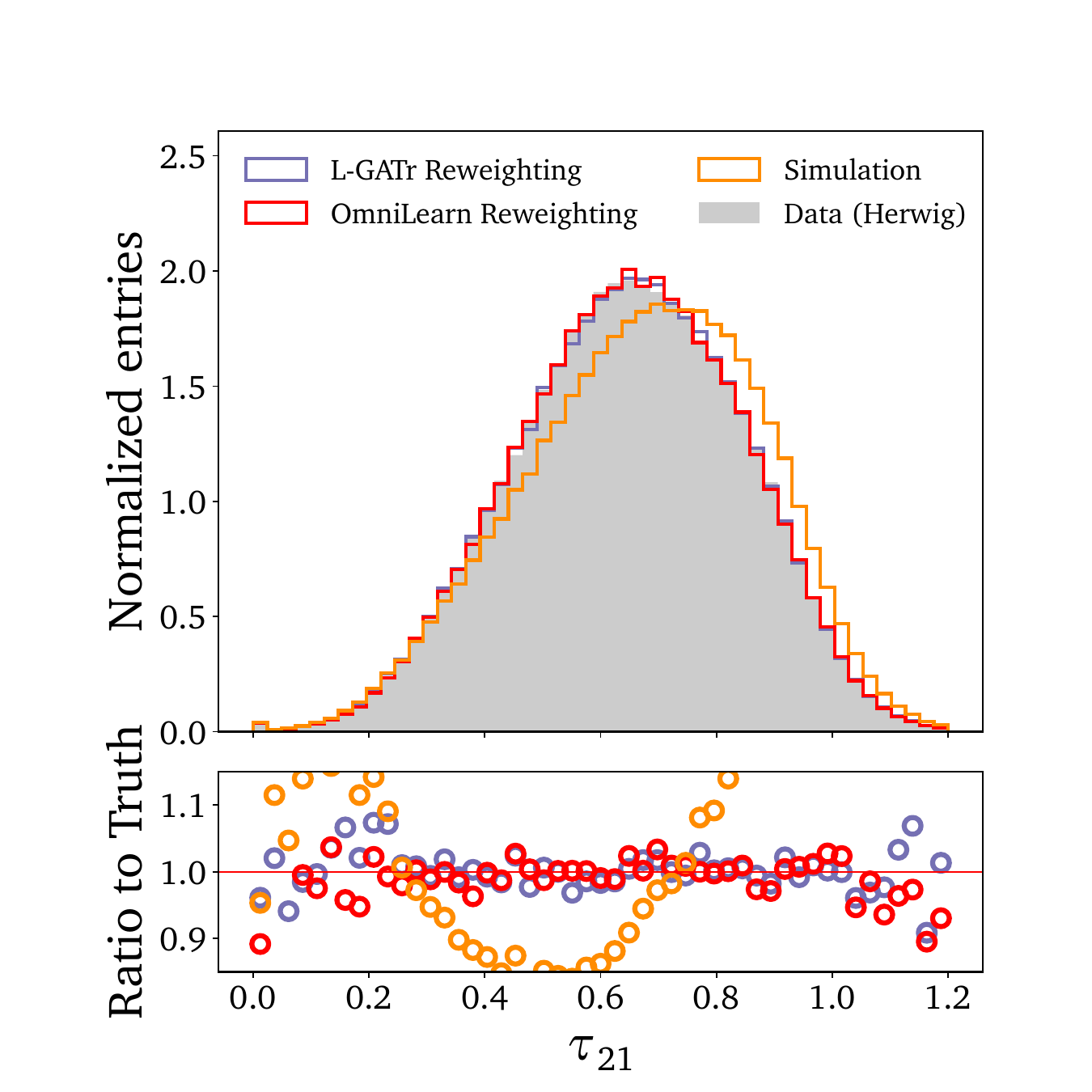} 
    \end{tabular}
    \caption{Reweighted distributions for six observables obtained from L-GATr and \textsc{OmniLearn} after 1 classifier training. The data events are generated with Herwig, and the simulation events are generated with Pythia. The L-GATr architecture consists of $10^6$ learnable parameters.}
    \label{fig:reweighted_results}
\end{figure}
%-------------------------------------------------

%-------------------------------------------------
\begin{table}
%\hspace{-0.55cm}
    \setlength{\tabcolsep}{4pt}
    \resizebox{\columnwidth}{!}{
    \begin{tabular}{l r cccccc}
        \toprule
        Method & Params. & $m$ & $w$ & $N$ & $\rm{log}\,\rho$ & $z_g$ & $\tau_{21}$ \\
        \midrule
        PET & $2\times 10^6$ & \val{0.10}{0.02} & \val{0.11}{0.02} & \bestval{0.15}{0.02} & \val{0.09}{0.01} & \val{0.05}{0.01} & \val{0.09}{0.01}\\
        \textsc{\textsc{OmniLearn}} & $2\times 10^6$  & \val{0.09}{0.02} & \val{0.08}{0.01} & \bestval{0.15}{0.02} & \bestval{0.06}{0.01} & \bestval{0.04}{0.01} & \bestval{0.07}{0.01}\\
        \midrule
        L-GATr-slim & $7\times 10^5$ &  \val{0.10}{0.02} & \val{0.10}{0.01} & \val{0.23}{0.03} & \val{0.08}{0.01} & \val{0.07}{0.01} & \val{0.08}{0.01} \\
        L-GATr & $10^6$ &  \bestval{0.08}{0.01} & \bestval{0.07}{0.01} & \val{0.16}{0.02} & \val{0.08}{0.01} & \val{0.06}{0.01} & \val{0.08}{0.01} \\
        L-GATr  & $2\times 10^6$ & \bestval{0.08}{0.01} & \bestval{0.07}{0.01} & \val{0.17}{0.02} & \val{0.09}{0.01} & \val{0.05}{0.01} & \bestval{0.07}{0.01} \\
        \bottomrule
    \end{tabular}}
    \caption{Detector-level triangular discriminators ($\times 1000$) between the Herwig and reweighted Pythia distributions after the first step of the first iteration of \textsc{OmniFold} using L-GATr, L-GATr-slim, PET trained from scratch and \textsc{OmniLearn} as event classifiers. The uncertainties are computed as the standard deviation of 100 uniform samples of each observable within their statistical uncertainties. The best results are in bold.}
    \label{tab:triangle-metrics-reco}
\end{table}
%-------------------------------------------------

%-------------------------------------------------
\begin{figure}[t]
    \hspace{-26pt}
    \setlength{\tabcolsep}{-5pt}
    \renewcommand{\arraystretch}{0}
    \begin{tabular}{@{}ccc@{}}
    \includegraphics[width=0.39\textwidth]{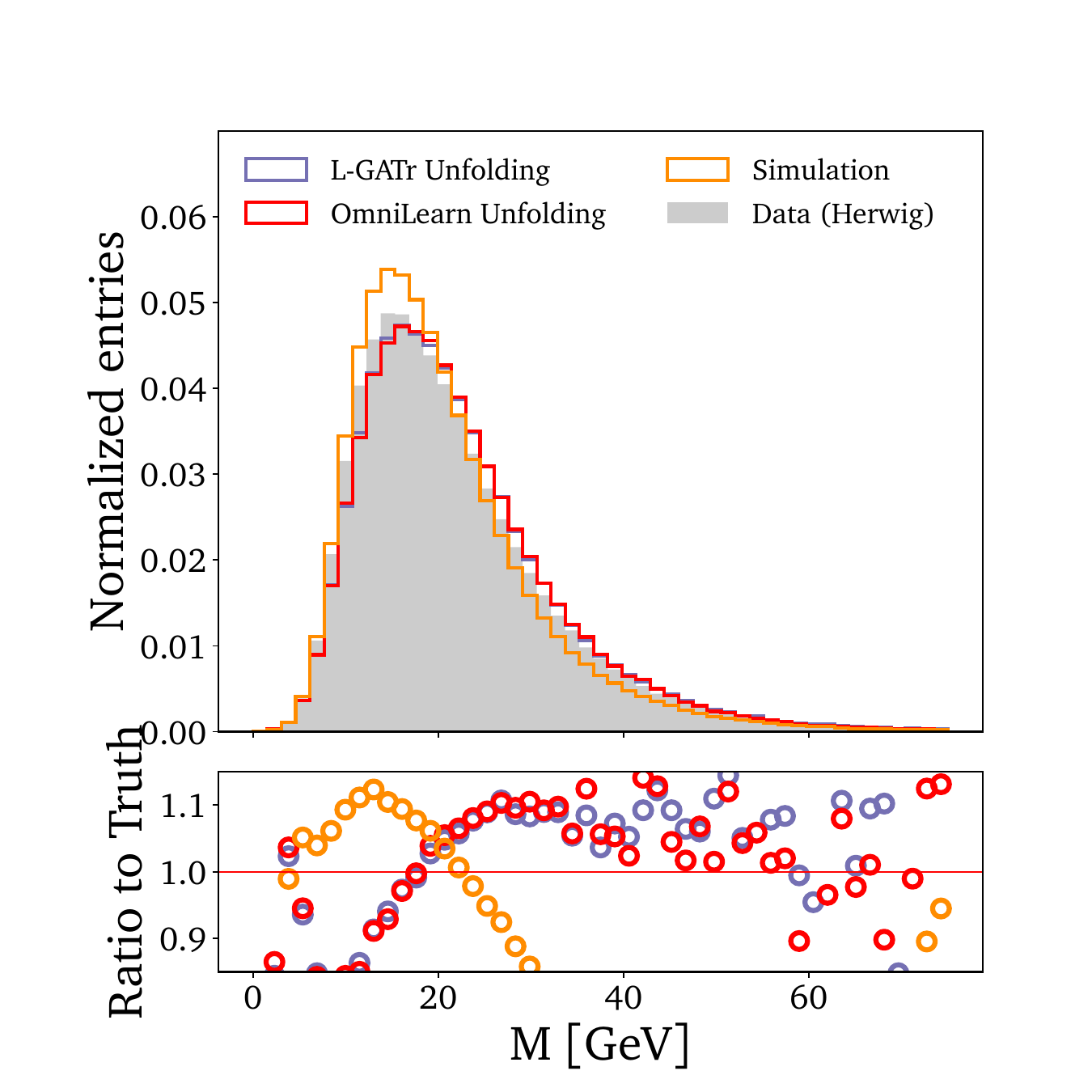}% &
    \hspace{-16pt}%
    \includegraphics[width=0.39\textwidth]{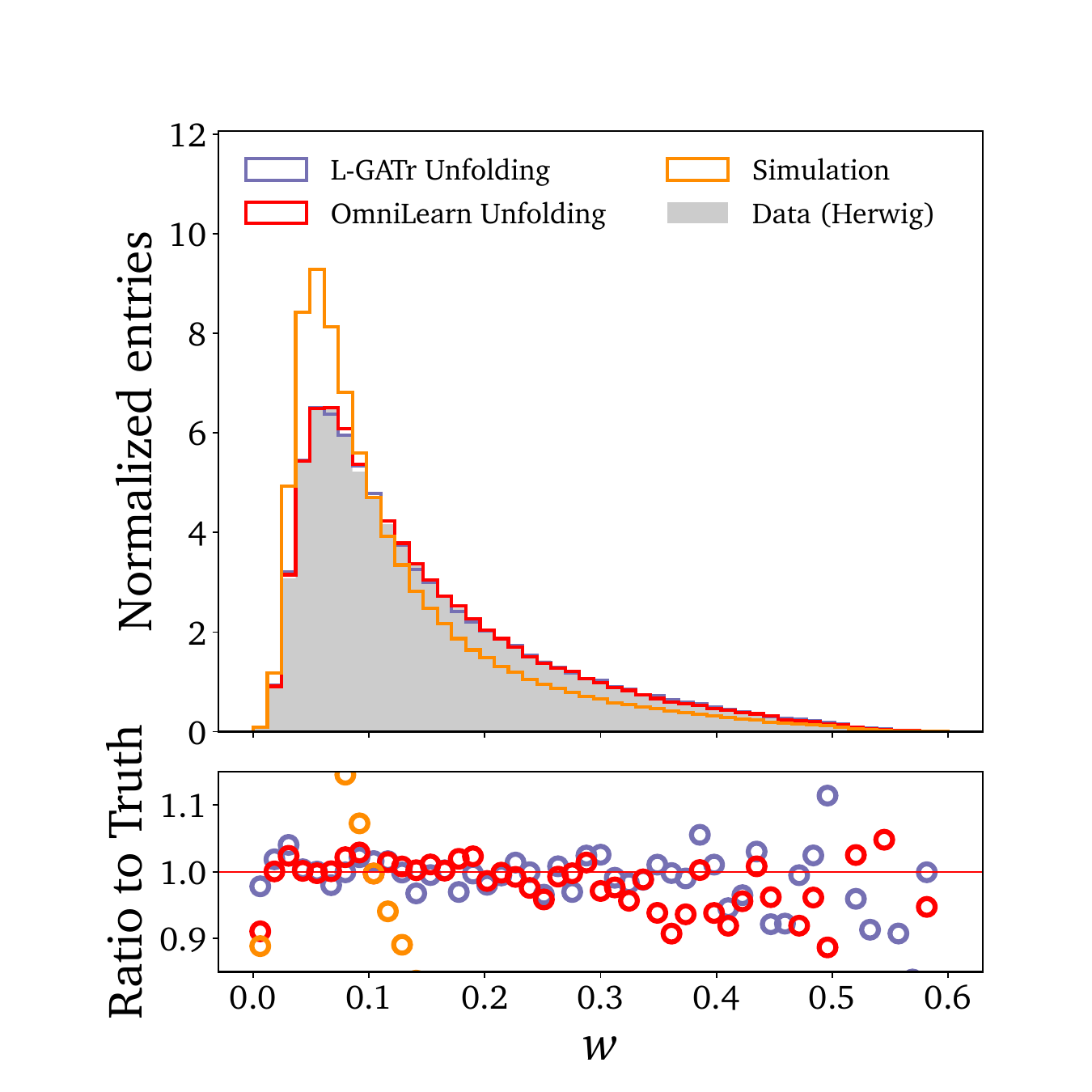}% &
    \hspace{-16pt}%
    \includegraphics[width=0.39\textwidth]{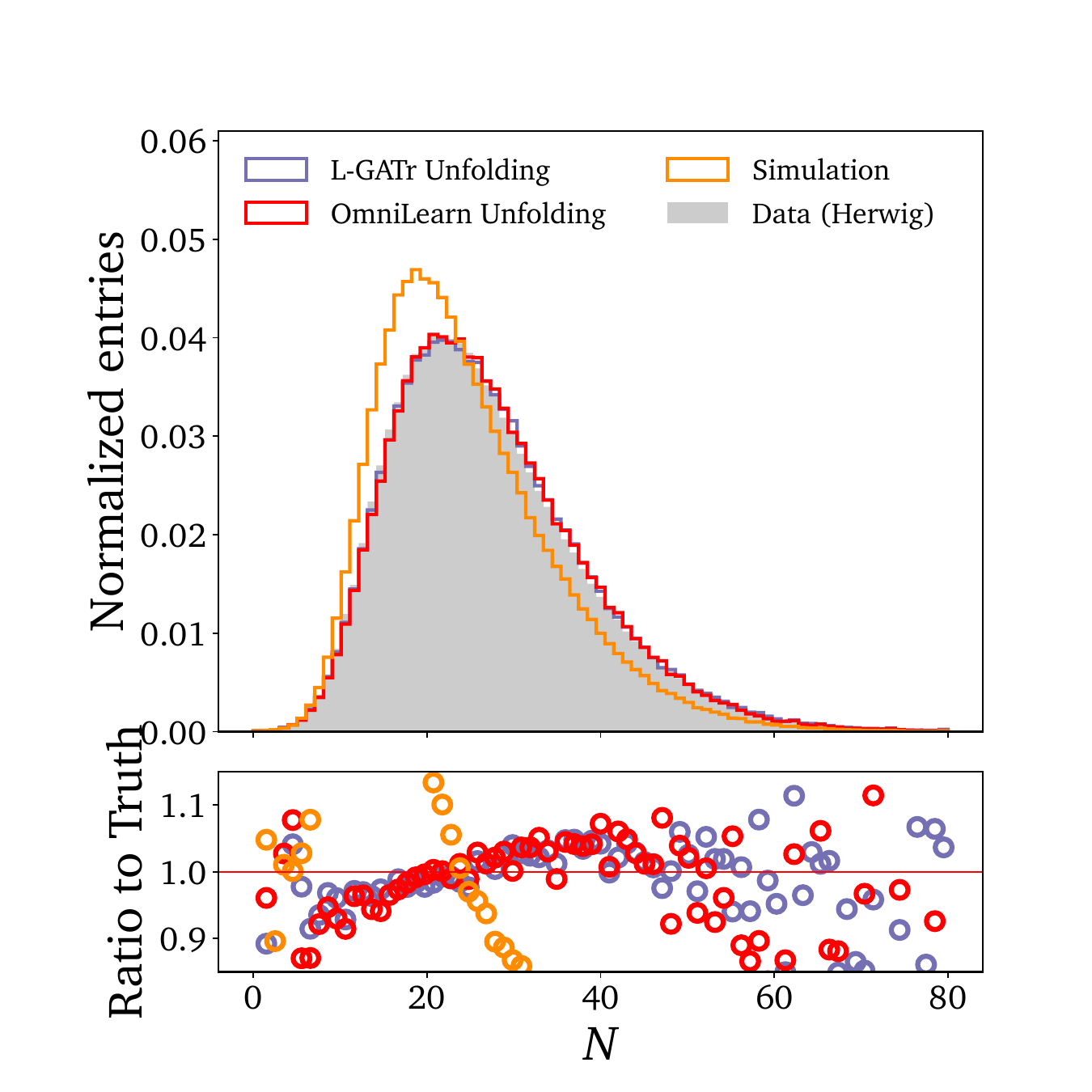}\\%[-2mm]
    \includegraphics[width=0.39\textwidth]{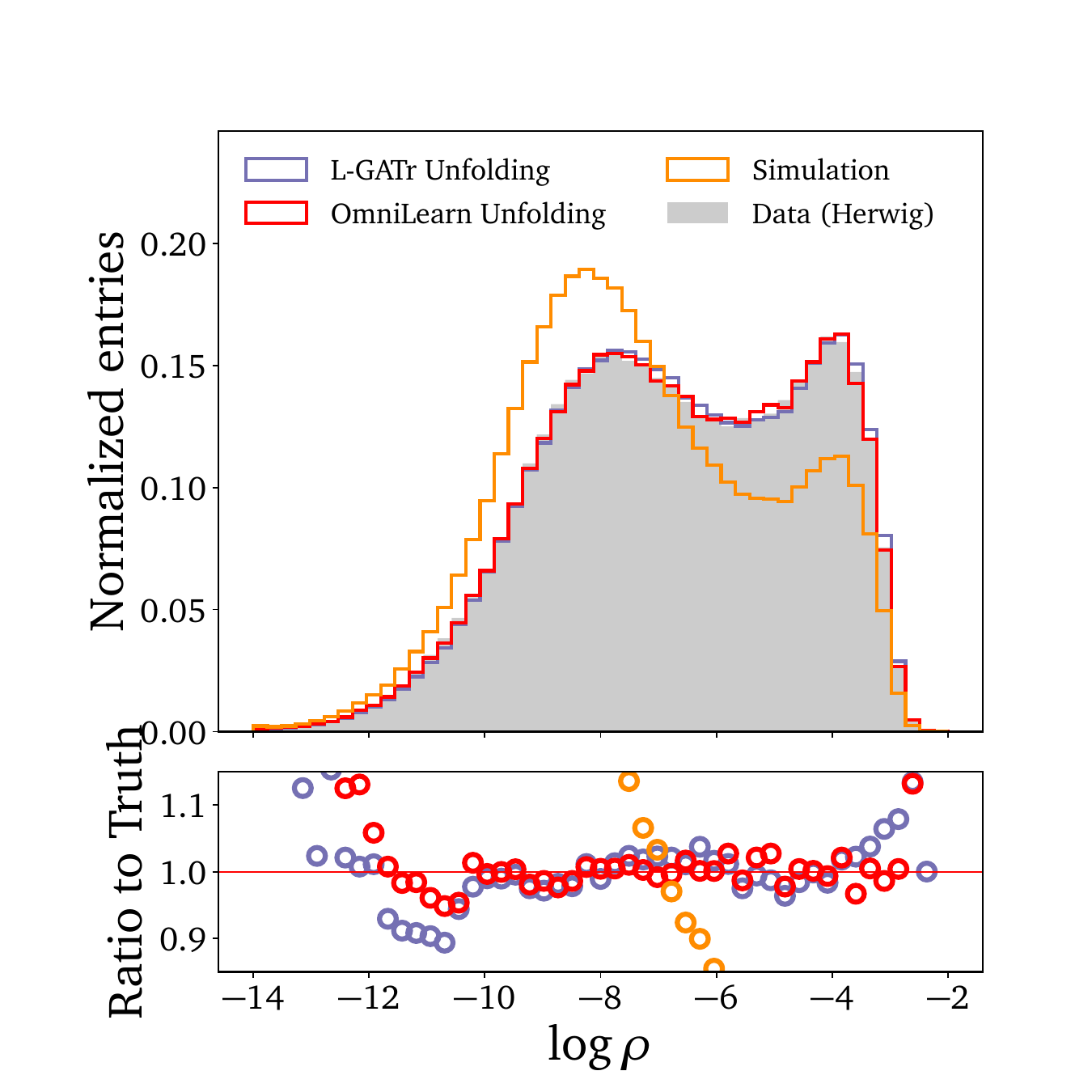}% &
    \hspace{-16pt}%
    \includegraphics[width=0.39\textwidth]{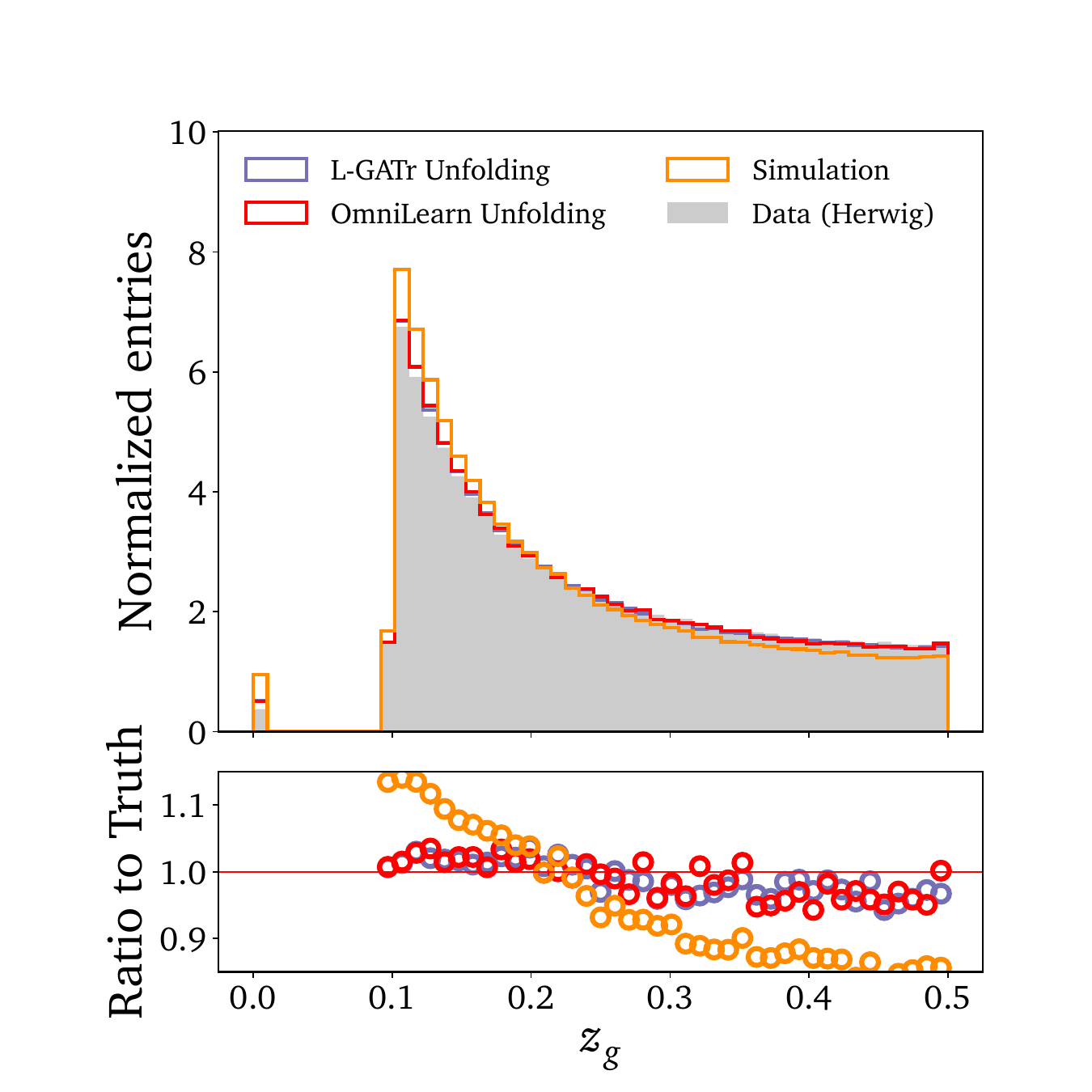}% &
    \hspace{-16pt}%
    \includegraphics[width=0.39\textwidth]{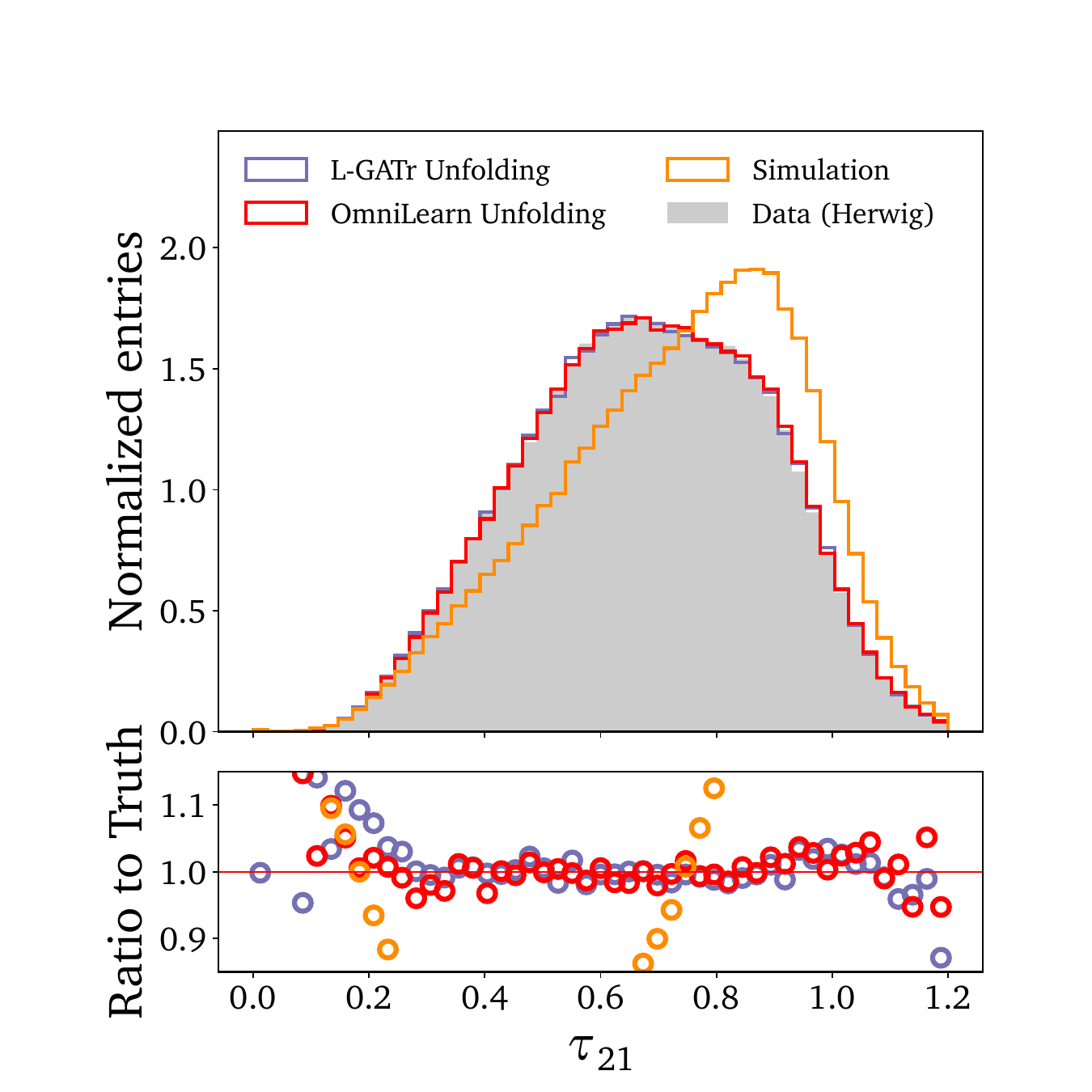}
    \end{tabular}
    \caption{Unfolded distributions for six observables  obtained from L-GATr and \textsc{OmniLearn} after 5 reweighting iterations. The data events are generated with Herwig, and the simulation events are generated with Pythia. The L-GATr architecture consists of $10^6$ learnable parameters.}
    \label{fig:unfolded_results}    
\end{figure}
%-------------------------------------------------

%-------------------------------------------------
\begin{table}
%\hspace{-0.55cm}
    \setlength{\tabcolsep}{4pt}
    \resizebox{\columnwidth}{!}{
    \begin{tabular}{l r c ccccc}
        \toprule
        Method & Params. & $m$ & $w$ & $N$ & $\rm{log}\,\rho$ & $z_g$ & $\tau_{21}$ \\
        \midrule
        PET & $2\times 10^6$ & \val{2.35}{0.06} & \val{0.22}{0.02} & \val{0.49}{0.03} & \val{0.17}{0.02} & \val{0.35}{0.02} & \val{0.12}{0.02}\\
        \textsc{\textsc{OmniLearn}} & $2\times 10^6$ & \val{2.06}{0.05} & \val{0.18}{0.02} & \val{0.33}{0.03} & \bestval{0.12}{0.01} & \bestval{0.27}{0.02} & \val{0.09}{0.01}\\
        \midrule
        L-GATr-slim & $7\times 10^5$ & \val{2.63}{0.07} & \val{0.14}{0.02} & \val{0.54}{0.04} & \val{0.14}{0.02} & \val{0.28}{0.02} & \bestval{0.07}{0.01} \\
        L-GATr & $10^6$ & \val{2.05}{0.06} & \bestval{0.12}{0.02} & \bestval{0.24}{0.02} & \val{0.23}{0.02} & \val{0.28}{0.02} & \val{0.08}{0.01} \\
        L-GATr & $2\times 10^6$ & \bestval{2.01}{0.06} & \val{0.17}{0.02} & \val{0.47}{0.04} & \val{0.13}{0.02} & \val{0.37}{0.03} & \val{0.08}{0.01} \\
        \bottomrule
    \end{tabular}}
    \caption{Particle-level triangular discriminators ($\times 1000$) between the Herwig and reweighted Pythia distributions after 5 iterations of \textsc{OmniFold} using the L-GATr, L-GATr-slim, PET and \textsc{OmniLearn} networks as classifiers. The uncertainties are computed as the standard deviation of 100 uniform samples of each observable within their statistical uncertainties. The best results are in bold.}
    \label{tab:triangle-metrics-hard}
\end{table}
%-------------------------------------------------
We use a standard set of observables to quantify the unfolding performance:
\begin{align}
    \begin{Bmatrix*}[r]
        {\text{jet mass}~m} & \phantom{xxx} {\text{$N$-subjettiness ratio}~\tau_{21}} &
        \phantom{xxx} {\text{groomed momentum fraction}~z_g} \\
        {\text{jet width}~w}& \phantom{xxx}{\text{constituent multiplicity}~N}&
        {\text{groomed mass}~\log \rho}
    \end{Bmatrix*}
\end{align}
We show histograms comparing the reweighted and target distributions at the detector level in Figure~\ref{fig:reweighted_results}. We also present the triangular discriminator distance metrics~\cite{850703,Gras:2017jty,Bright-Thonney:2018mxq} in Table~\ref{tab:triangle-metrics-reco}, which are computed as
\begin{equation}
    \Delta_{\lambda}(p,q) = \frac{1}{2} \sum_i\frac{(p_i(\lambda) - q_i(\lambda))^2}{p_i(\lambda) + q_i(\lambda)} \, (\times 10^{-3}),
\end{equation}
where $\lambda$ denotes the examined observable, $p$, $q$ are the two compared distributions represented as histograms and the sum is done over all bins.

%: jet mass $m$, jet width $w$, jet multiplicity $N$, groomed mass $\rm{log}\, \rho$, groomed momentum fraction $z_g$, and N-subjettiness ratio $\tau_{21}$. 
In general, we observe a significant improvement from \textsc{OmniLearn}, L-GATr, and L-GATr-slim compared to the PET trained from scratch. This illustrates that explicit and implicit physics priors offer a clear performance boost. The L-GATr and \textsc{OmniLearn} results are similar for all metrics, with a small edge for \textsc{OmniLearn} over L-GATr for this reweighting step. There is  no clear improvement in the L-GATr results from including the scalar \textsc{OmniLearn} inputs, increasing the network size beyond $10^6$ parameters, or prolonged training. This suggests that the L-GATr performance is limited by the amount of training data. However, we also find that it is possible to match \textsc{OmniLearn} by using a smaller network size and a simpler set of data inputs. 

The second \textsc{OmniFold} step is classifier reweighting at particle level. The final result of this procedure after five iterations is the actual unfolding. We show the same set of histograms as before, but now for particle-level reweighting in Figure~\ref{fig:unfolded_results}, and the triangular discriminator distance metrics in Table~\ref{tab:triangle-metrics-hard}. Again, the explicit and implicit physics priors lead to a significant performance boost, now with a small edge for L-GATr over \textsc{OmniLearn}. As before, the L-GATr performance does not change when upscaling the network or the training time, indicating the limiting effect of the training dataset size.

Finally, we observe that L-GATr-slim underperforms with respect to L-GATr on the majority of metrics by a small margin, but it reaches comparable and even superior performance to the full model for several observables. This suggests that the significantly faster L-GATr-slim implementation can indeed be a viable alternative for L-GATr in scenarios with high computational constraints~\cite{Petitjean:2025zjf}. 

Let us remark that the differences in performance that we observe for the L-GATr networks upon retraining are substantial enough that for some singular trainings we observe that the hierarchy of results quality is completely reordered. We have no clear explanation as to why the results from the L-GATr networks are so sensitive to the weight initialization seeds. Since we only observe this behavior on this task, we can conjecture that this is caused by the gap between the training target and the examined metrics. More specifically, it is possible that the training loss can be optimized in multiple ways such that the correct modeling of a different set of observables is promoted in each instance.

%%%%%%%%%%%%%%%%%%%%%%%%%%%%%%%%%%%%%%%%%%%%%%%%%%%%%%%%%%%%
\subsection{Likelihood ratio estimation for $ep$ collisions}
\label{h1}
\label{h1}

As a second example, we re-weight high Q$^{2}$ deep-inelastic scattering (DIS) events at the H1 detector, i.e. with a distinct collision type, detector, and event generator. The simulated data is the same as in Refs.~\cite{Arratia:2021tsq,Arratia:2022wny,Long:2023mrj}. We use particle-level simulations from Djangoh 1.4~\cite{Charchula:1994kf}  and Rapgap 3.1~\cite{Jung:1993gf}, where Djangoh plays the role of `data' and Rapgap the role of `simulation'. The particle-level events are reconstructed using an energy-flow algorithm~\cite{energyflowthesis,energyflowthesis2,energyflowthesis3} after an H1 detector simulation with \textsc{Geant}3~\cite{Brun:1987ma}. They are required to have $Q^2 > 100$ GeV$^2$ and at least one particle in the event. Detector-level events are required to have $Q^{2}> 150$ GeV$^{2}$, inelasticity $0.08<y<0.7$, and particles with $p^\text{part}_\text{T} >$ 0.1 GeV and $-1.5 < \eta^\text{part.} < 2.75$. Both Rapgap and Djangoh simulations have 30 constituents per jet and amount to a total of $2.5\times 10^6$ events for training, $4\times 10^6$ events for testing, and $5\times 10^5$ events for validation. 

We again classify jets produced by one generator from the other. Compared to Section~\ref{zjets}, the differences between the constituent-level and jet-level features from the two generators are much more subtle. We again use the features listed in Table~\ref{tab:zjets_inputs} as inputs to L-GATr and handle the information at the constituent and jet levels following Section~\ref{lgatr}. 
%For both constituents and jets, the four-momenta are provided in the (E, $p_{x}$, $p_{y}$, $p_{z}$) representation, as required by the L-GATr architecture. 
We standardize the scalar features prior to training.
%\mpar{Only here or also in 3.1? vb: Yes, already updated the text to mention it.} 

\begin{figure}[h!]
    \centering
    \setlength{\tabcolsep}{3pt}
    \begin{tabular}{cc}
    \includegraphics[width=0.39\textwidth]{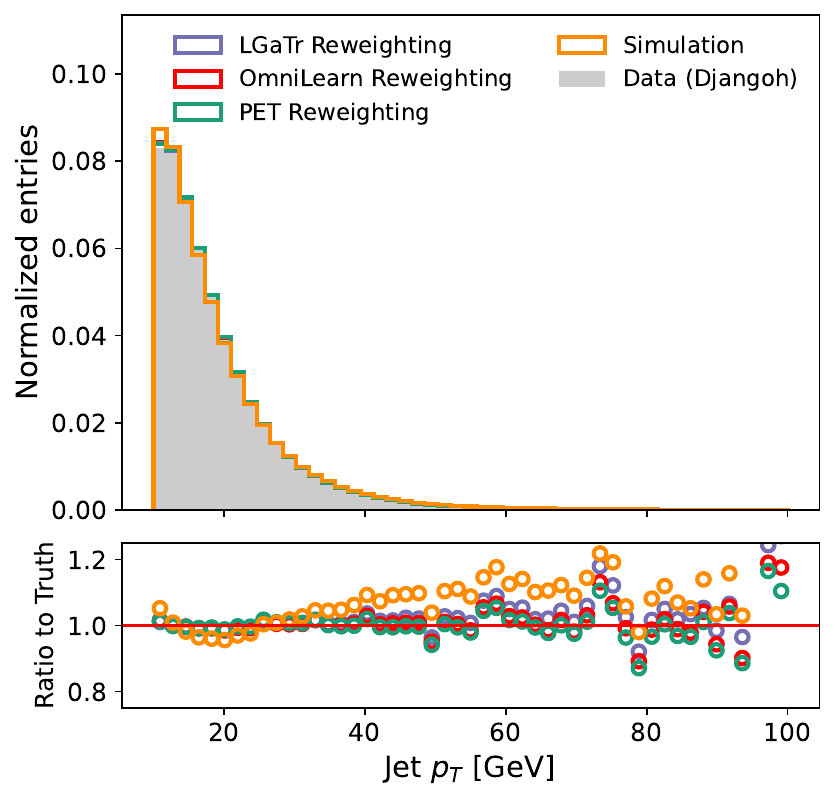} &
    \includegraphics[width=0.39\textwidth]{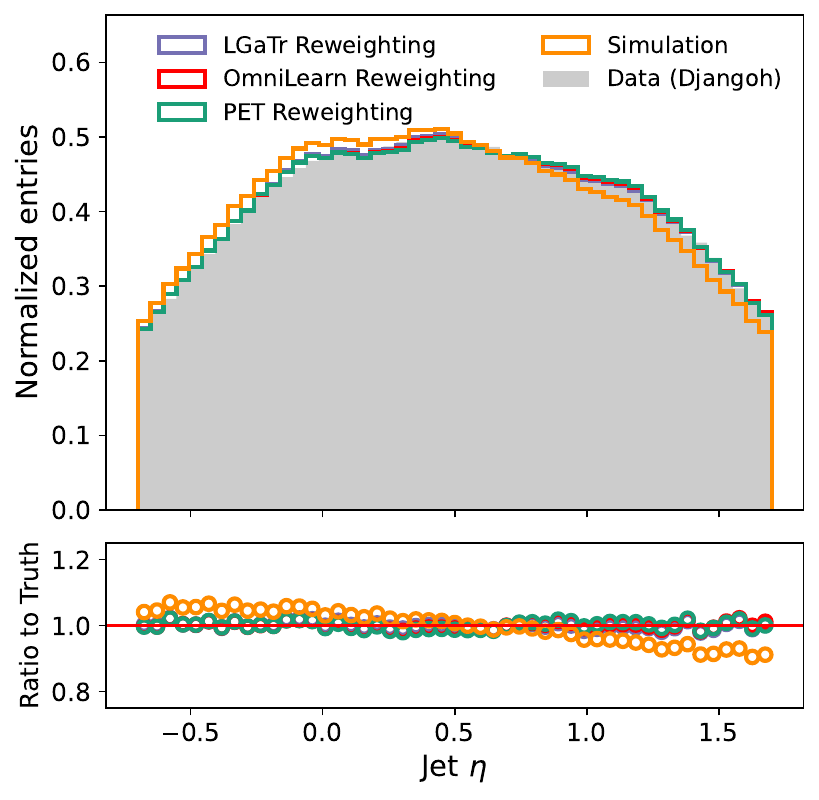} \\
    \includegraphics[width=0.39\textwidth]{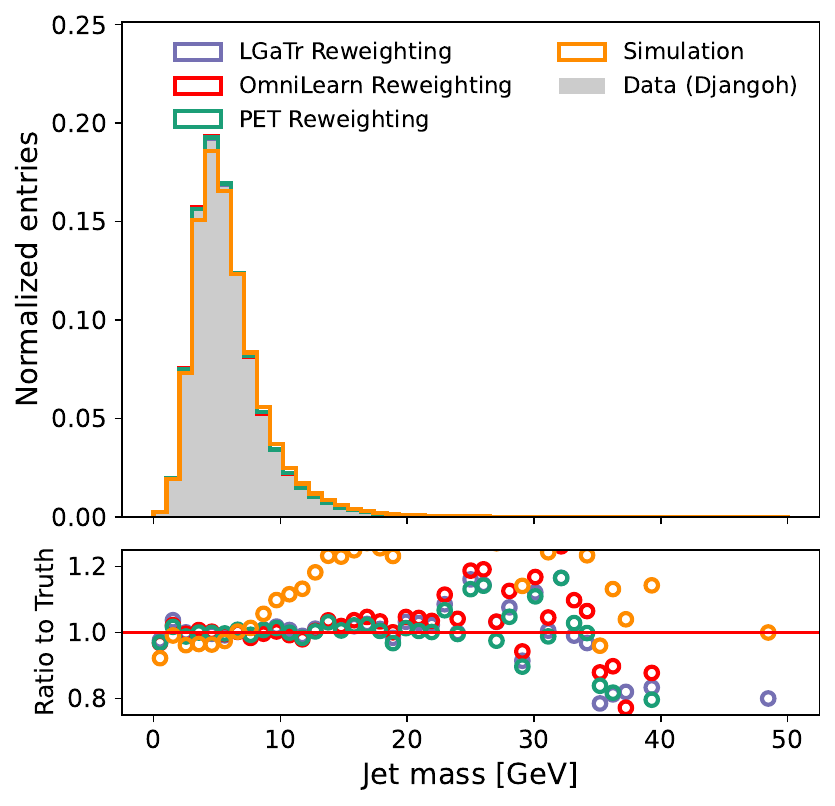} &
    \includegraphics[width=0.39\textwidth]{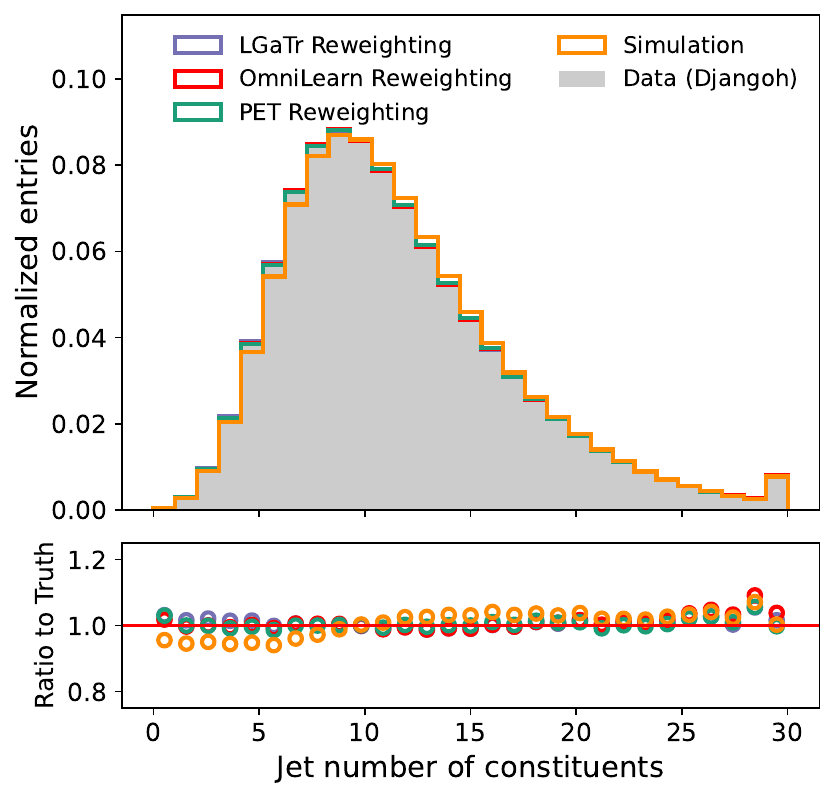}
    \end{tabular}
    \caption{Reweighted distributions for four observables obtained from L-GATr, PET and \textsc{OmniLearn}. The data events are generated with Djangoh, and the simulation events are generated with Rapgap. The L-GATr architecture consists of $10^6$ learnable parameters.}
    \label{fig:reweighted_results_h1}
\end{figure}

As in Section~\ref{zjets}, we evaluate two L-GATr builds with approximately $10^6$ and $2 \times 10^6$ parameters trained from scratch and compare their performance to \textsc{OmniLearn}. 
%L-GATr networks with the same sizes as Section~\ref{zjets} are trained.
To match the \textsc{OmniLearn} training setup as closely as possible, we use a batch size of 256, an initial learning rate of $3\times 10^{-5}$, and a weight decay of 0.2. A scan over batch size, learning rate, and weight decay is discussed in detail in Appendix~\ref{hyperparameters_h1}. All models are trained for 20 epochs, with no improvement in validation loss observed beyond epoch 10. The resulting reweighted distributions are shown in Figure~\ref{fig:reweighted_results_h1}. All three methods successfully correct the Rapgap simulation towards the Djangoh reference across all four observables, with the reweighted distributions lying significantly closer to the data than the unweighted simulation. L-GATr, \textsc{OmniLearn} and PET show comparable performance and achieve good agreement with the data across the full kinematic range, with slightly larger deviations from the reference in the tails of the distributions. 

%-------------------------------------------------
%-------------------------------------------------

%-------------------------------------------------
\begin{table}[H]
%\hspace{-0.8cm}
\small
\centering
\begin{tabular}{lccc}
\toprule
 Method
 & AUC & Acc & $1/\epsilon_B \;(\epsilon_S = 0.5)$ \\
\midrule

PET      & 0.5691 & 0.547 & $2.467 \pm 0.002$ \\
\textsc{OmniLearn}  & 0.5695 & 0.547 & $2.470 \pm 0.003$ \\
L-GATr ($10^6$ parameters)           & 0.5603  & 0.541 & $2.396 \pm 0.003$ \\
L-GATr ($2 \times 10^6$ parameters)           & 0.5630  & 0.543 & $2.408 \pm 0.002$ \\

\bottomrule
\end{tabular}
\caption{Performance comparison of classifiers on the H1 dataset.}
\label{tab:classifier_performance}
\end{table}
%-------------------------------------------------

The network performance is evaluated using the accuracy, AUC and rejection rate at 50\% signal efficiency on the test dataset, and is listed in Table~\ref{tab:classifier_performance}. For robustness, each model is trained three times with different random seeds, and the best performing run is reported. We find that L-GATr with $10^6$ parameters is slightly, but consistently worse than, both, \textsc{OmniLearn} and the PET trained from scratch. Given that \textsc{OmniLearn} only leads to a small improvement over a simple PET training, it appears that the architecture itself is the main performance driver for this task. The cause for this may be the local feature treatment present in PET, which is the main architecture feature that is completely absent in L-GATr. Increasing the L-GATr size to $2 \times 10^6$ parameters, our closest match to \textsc{OmniLearn}, yields only marginal improvements. In particular, the limited improvement when increasing the L-GATr capacity suggests that scaling the number of model parameters does not match the performance gains provided by the design of the PET architecture and to a lesser degree \textsc{OmniLearn}’s large-scale pretraining on the data side.

%%%%%%%%%%%%%%%%%%%%%%%%%%%%%%%%%%%%%%%%%%%%%%%%%%%%%%%%%%%%
\subsection{Weakly supervised anomaly detection}
\label{ad}
Anomaly detection employs ML-methods to conduct model-independent searches for new phenomena in particle physics~\cite{Heimel:2018mkt,Nachman:2020lpy,Kasieczka:2021xcg} without a given signal hypothesis. One class of techniques relies on Classification Without Labels (CWoLa), where a classifier is trained to separate a background-only reference dataset from the data, which is a mixture of many background events and a fraction of signal events~\cite{Metodiev:2017vrx}. The signal fraction is small by assumption, so the background-only reference dataset and the data are assumed to be nearly identical, making this type of anomaly detection another example of a classification problem with nearly identical classes.

For this weakly supervised anomaly detection, we have to construct a suitable background template using data-driven techniques~\cite{Hallin:2021wme}. 
We compare the performance of L-GATr and \textsc{OmniLearn} classifiers on the LHC Olympics anomaly detection benchmark dataset~\cite{LHCOlympics}, solving the weakly supervised classification task with background events sampled directly from a predefined distribution. This is often termed `idealized anomaly detection', since in real applications such samples are not available. 

Our background consists of di-jet final states produced through QCD processes, while the signal is represented by resonant boson production following the decay chain
\begin{align}
& A \to B(\to q q^{\prime})\, C(\to q q^{\prime})  \notag \\
& \qquad \text{with}  \qquad m_{A} = 3.5~\text{TeV}, m_{B} = 0.5~\text{TeV}, m_{C} = 0.1~\text{TeV} \; .
\end{align}
Following in particular Ref.~\cite{Mikuni:2025wjk}, we define a signal where the di-jet mass satisfies
\begin{align}
 3.3~\text{TeV} < m_{jj} < 3.7~\text{TeV} \; .
\end{align}
Events in this region are used to construct two datasets that form the stand-ins for the data and the background model. The data consists of $10^5$ background events mixed with a varying number of signal events. The background model is composed of 350k independently generated background events.

Given that both L-GATr and \textsc{OmniLearn} are large point-cloud models, we consider full phase-space anomaly detection~\cite{Buhmann:2023acn}, where the constituent kinematics of each jet are provided to the classifiers, in addition to jet-level and event-level kinematic information. 

%-------------------------------------------------
\begin{table}[b!]
%\hspace{-0.8cm}
\small
\centering
\begin{tabular}{c c c c c}
\toprule
Input Type & Scalar & 4-vector & One-hot & Event \\
\midrule
Jet constituents &
\makecell{
$\Delta y_{i,\text{jet}}$, $\Delta \phi_{i,\text{jet}}$, $\log p_\textrm{T}$, $\log E$, \\
$\log \frac{p_{\textrm{T},i}}{p_{\textrm{T},\text{jet}}}$, $\log \frac{E_i}{E_{\text{jet}}}$, $\Delta R_{i,\text{jet}}$
}
& $E$, $p_x$, $p_y$, $p_z$ & $\mathbbm{1}_{j1}$, $\mathbbm{1}_{j2}$ & $m_{jj}$ \\[7mm]
Jets &
$p_\textrm{T}$, $y$, $\phi$, $E$, $m$
& $E$, $p_x$, $p_y$, $p_z$ & $\mathbbm{1}_{j1}$, $\mathbbm{1}_{j2}$ & $m_{jj}$ \\
\bottomrule
\end{tabular}
\caption{Input features to the L-GATr networks for the LHCO data set. Each feature is provided for each item in the sequence representing the event, i.e. two jets and a variable number of jet-constituents. For constituents, jet-level features are zeroed, and for jets, constituent-level features are zeroed. The one-hot features are set to one if the item corresponds to the higher or lower $p_\textrm{T}$ jet, respectively.}
\label{tab:lhco_inputs}
\end{table}
%-------------------------------------------------

We train an L-GATr model with $1.8 \times 10^6$ trainable parameters and compare to the \textsc{OmniLearn} performance from Ref.~\cite{Mikuni:2025wjk}. In contrast to Sections~\ref{zjets} and~\ref{h1}, full phase space anomaly detection requires processing events rather than individual jets. We allow L-GATr to process events by providing both scalar and 4-vector representations of the constituents of all jets in each event. The constituents are identified with the appropriate jet through a one-hot encoding added as a scalar feature. The jet-level kinematics are provided as an additional token following Section~\ref{lgatr}. The scalar inputs used to represent the constituents and jets are identical to Refs.~\cite{Buhmann:2023acn, Mikuni:2025wjk}. All scalar input features are standardized. Finally, the di-jet mass $m_{jj}$, known to be  particularly useful, is included as an extra scalar feature for each item in the event sequence. A summary of the inputs given to L-GATr is provided in Table~\ref{tab:lhco_inputs}. A full description of the hyperparameters is provided in Appendix~\ref{hyperparameters_ad}.

%-------------------------------------------------
\begin{figure}[t]
    \centering
    \includegraphics[width=0.5\linewidth]{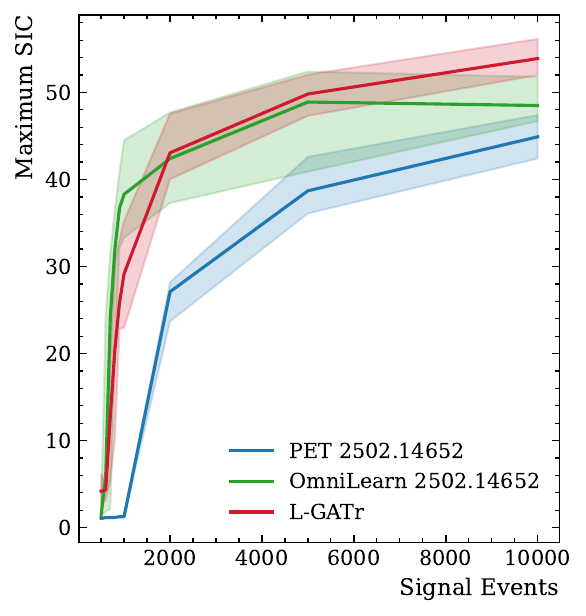}
    \caption{Maximum SIC against the number of injected signal events for the randomly initialized PET and L-GATr models, as well as the pretrained \textsc{OmniLearn}. Solid lines plot the mean maximum SIC across an ensemble of 10 independent trainings, and shaded bands represent the 68\% confidence band over the ensemble.}
    \label{fig:ad_max_sics}
\end{figure}
%-------------------------------------------------

Following standard practice in the anomaly detection literature, we use the maximum of the significance improvement characteristic curve (SIC) as a performance metric. We train L-GATr classifiers with 500 to $10^4$ injected signal events and then take the model checkpoint that minimizes the validation loss to calculate the maximum SIC on a held-out testing set. For each signal injection, we train ten classifiers and evaluate the maximum SIC produced by each. We then calculate the mean maximum SIC over the ensemble.

The performance of L-GATr and the different reference benchmarks as a function of the number of injected signal events is shown in Figure~\ref{fig:ad_max_sics}. Shaded bands represent the 68\% confidence band over the 10 independent classifier trainings. The results are compared to the randomly initialized PET model and \textsc{OmniLearn}~\cite{Mikuni:2025wjk}. The physics-primed L-GATr and \textsc{OmniLearn} substantially outperform the PET model in terms of the raw classification power captured by the maximum SIC performance metric.

An application of these classifiers to a realistic data analysis would require a more careful consideration of potential false positives~\cite{Gambhir:2025afb}. The maximum SIC of L-GATr and \textsc{\textsc{OmniLearn}} are the same within uncertainties for all signal injections tested. L-GATr and L-GATr-slim models with an order of magnitude fewer trainable parameters showed a significantly reduced performance, indicating that this event-level classification requires larger networks even when explicit physics priors are used.

\clearpage
%%%%%%%%%%%%%%%%%%%%%%%%%%%%%%%%%%%%%%%%%%%%%%%%%%%%%%%%%%%%
\section{Outlook}
\label{conclusion}

We have presented a systematic comparison between two strategies for incorporating known physics structures into neural networks: explicit Lorentz equivariance, represented by L-GATr, and implicit priors learned through large-scale pretraining, represented by \textsc{OmniLearn}. We considered precision classification tasks, involving nearly identical classes:
%as the performance of L-GATr and \textsc{OmniLearn} in classification tasks with significant class differences has already been studied in detail.
%
\begin{enumerate}
\item As reweighting classifiers in \textsc{OmniFold}, the explicit and implicit approaches show very similar performance with no clear pattern for a preference. Both estimate the likelihood ratio within the accuracy supported by the dataset. Even the much more efficient L-GATr-slim reaches competitive performance on some observables. 

\item For the likelihood ratio estimation between DIS events, L-GATr consistently underperforms, even with increased network size. This suggests that the efficiency gain from assuming broken Lorentz-equivariance does not match the benefits from local feature processing and large-scale pre-training for the given size and complexity of the H1 training dataset. 

\item For weakly supervised anomaly detection, L-GATr and \textsc{OmniLearn} achieve comparable performance. We see slight potential benefits of the implicit approach for small signal injections and of the explicit approach for large signal injections.
\end{enumerate}
%L-GATr's performance in this setting suggests that explicitly enforcing Lorentz symmetry can be especially effective in classification problems involving nearly identical classes, where subtle kinematic correlations are critical. 
%The observed underperformance of the smaller L-GATr network demonstrates that explicit physics priors alone are not sufficient to achieve optimal performance without adequate model capacity in the context of full phase-space classification tasks. %While equivariance improves data efficiency, achieving optimal performance still requires a minimum representational complexity, particularly in 
%
Finally, it is not clear whether the two approaches learn exactly the same underlying physics. One possible way to bring more clarity into this matter would be a dedicated study into the internal representations of pretrained and equivariant networks. However, such an analysis would be very challenging, since it would entail probing data representations with very high dimensions and there is no clear indication about which features of either the data or the architectures are relevant to improve performance on any given task. Furthermore, there is no practical impediment for combining the two approaches. %approaches by applying the pretraining strategy on an equivariant network backbone. In fact, this unified approach can result in marginal performance gains on complex tasks~\cite{Brehmer:2024yqw,Spinner:2025prg}. 
Potential benefits of leveraging the two strategies at the same time will again have to be evaluated in terms of performance, computational cost, and usability in practice.
%. We expect that the results presented on this study will be useful to determine the utility of such a combined approach for different tasks in collider physics.   

\section*{Code and Data availability}
The code for this work is available at \url{https://github.com/stanford-ai4physics/physics-priors}. The L-GATr architectures are implemented using version 1.4.3 of the package available at \url{https://github.com/heidelberg-hepml/lgatr}.

\section*{Acknowledgments}
We thank Jonas Spinner for expert advice on the tuning of the L-GATr-slim architecture and for many helpful discussions. BN is supported
by the U.S. Department of Energy (DOE), Office of Science under contract DE-AC02-76SF00515. KG and DW are supported by the DOE Office of Science. TW is supported by the National Science Foundation under Grant No. 2311666. This research
used resources of the National Energy Research Scientific Computing Center, a DOE Office of Science User
Facility supported by the Office of Science of the DOE under Contract No. DE-AC02-05CH11231 using NERSC award HEP-ERCAP0035546. VBP and TP are supported by the Deutsche Forschungsgemeinschaft (DFG, German Research Foundation) under grant
396021762 – TRR 257 Particle Physics Phenomenology after the Higgs Discovery. VBP acknowledges financial support from the Real Colegio Complutense at Harvard University Postdoctoral Research Fellowship. VM is supported by JST EXPERT-J, Japan Grant Number JPMJEX2509. This work benefited from interactions with the \href{http://iaifi.org/}{NSF AI Institute for Artificial Intelligence and Fundamental Interactions}, which is supported by the National Science Foundation under Cooperative Agreement PHY-2019786.
We express our thanks to all those involved in securing not only the H1 data but also the software and working environment for long term use, allowing the unique H1 data set to continue to be explored. The transfer from experiment specific to central resources with long term support, including both storage and batch systems, has also been crucial to this enterprise. We therefore also acknowledge the role played by DESY-IT and all people involved during this transition and their future role in the years to come.

\appendix
\clearpage
%%%%%%%%%%%%%%%%%%%%%%%%%%%%%%%%%%%%%%%%%%%%%%%%%%%%%%%%%%%%
\section{Computational resource analysis}
\label{compute}
We compare the computational requirements for training L-GATr, L-GATr-slim and \textsc{OmniLearn} on a benchmark classification task.
% The most obvious way to quantify the computational resources spent by a network training is by looking at the training time and the peak memory usage. However, these quantities are highly dependent on the computing setup used for the training. Factors such as the training batch size, the distribution of the computation across multiple GPUs and the data loading routines can offer a biased picture of the actual performance of a neural network.
Naive comparisons using training time are highly dependent on the compute resources, batch size, data loading routines, and other factors.
In order to benchmark the usage of training resources properly, we perform our measurements in a fixed environment with a single Nvidia A100 GPU and floating point precision to \verb|float32|. 
We measure the forward pass time, peak memory usage and floating point operations (FLOPs), as well as the compute required to train both networks.
The full training runs last for 20 epochs, which is a representative training length for all the tasks considered in this paper.  

We perform these measurements by training both networks on a dataset consisting of $2\times 10^6$ jet events with 101 particles on a single A100 GPU. The 4-momenta of each jet constituent are used as inputs for L-GATr and L-GATr-slim, while 13 features per particle and 4 features per jet are used for \textsc{OmniLearn}. This setup serves as a proxy for the $Z$+jets dataset studied in Section~\ref{zjets}. We also consider the same network sizes studied in that section. We consider a realistic batch size of 512 for the full training and a batch size of 1 for the forward pass measurements. In all cases, we set the numerical floating point precision to \verb|float32|. 

We use the \texttt{time.perf\_counter()} module to measure time and average over 10 independent measurements. We measure the peak memories by using 
\texttt{torch.cuda.\allowbreak max\_memory\_allocated} 
in the case of L-GATr and L-GATr-slim and \texttt{tf.config.\allowbreak experimental.\allowbreak get\_memory\_info} in the case of \textsc{OmniLearn}. Finally, we measure the compute FLOPs using model profilers from the respective libraries for each network. The floating point operations required by a full training are calculated by multiplying the operations required by a forward pass by $3\times \text{dataset size} \times \text{epochs}$, where the 3 coefficient comes from the added computation cost involved in the backward pass.

\begin{table}
%\hspace{-0.8cm}
\small
\centering
\begin{tabular}{lccc}
\toprule
 & L-GATr & L-GATr-slim & \textsc{OmniLearn} \\
\midrule

Forward pass time (ms) 
& $58.16 \pm 0.45$ 
& $15.39 \pm 0.16$ 
& $191.22 \pm 2.04$ \\

\midrule

Forward pass compute (FLOPs) 
& $3.7 \times 10^{9}$ 
& $2.4 \times 10^{8}$ 
& $4.8 \times 10^{8}$ \\ 

Full training compute (FLOPs) 
& $4.4 \times 10^{17}$ 
& $2.9 \times 10^{16}$ 
& $5.8 \times 10^{16}\,(+1.2\times 10^{19})$ \\

\midrule

Peak memory (MB) 
& $154.41$ 
& $47.95$ 
& $22.26$ \\ 

\midrule

Number of parameters 
& $10^{6}$ 
& $7 \times 10^{5}$ 
& $2 \times 10^{6}$ \\

\bottomrule
\end{tabular}
\caption{Performance comparison between L-GATr, L-GATr-slim and \textsc{OmniLearn} in terms of training speed, peak memory, and compute per forward pass on a single A100 GPU using a single jet event with 101 particles. Forward-pass time is averaged over 10 independent measurements. We also report the total compute over a full training of 20 epochs. The additional computation cost from the pretraining of \textsc{OmniLearn} is shown in parentheses.}
\label{tab:performance_comparison}
\end{table}

We present the results of our measurements in Table~\ref{tab:performance_comparison}. L-GATr requires roughly 10 times more compute and memory than \textsc{OmniLearn} over a single forward pass. This is an expected property of L-GATr. However, L-GATr is 3 times faster than \textsc{OmniLearn} for a single forward pass. We attribute this to differences on the backend implementation of the two networks. More specifically, L-GATr is built on a suite of attention backends that optimize network speed and memory usage. Once we consider L-GATr-slim, we observe that the compute cost and memory usage improve by a factor of 10 and 3 respectively. Additionally, the network becomes even faster, improving by a factor of 4. These results also represent a significant improvement with respect to \textsc{OmniLearn}, positioning L-GATr-slim as our most efficient framework.

We also consider the FLOPs spent during the pre-training of \textsc{OmniLearn} to our full training compute measurements. This element marks an important difference with respect to L-GATr, which is always performed from scratch. Once we add it to the total cost, we observe that \textsc{OmniLearn} requires roughly 25 times more computing resources than L-GATr. This cost is not present for ordinary fine-tunings that use the readily available pre-trained weights, but it should be taken into account in cases where a new pre-training is required.

%%%%%%%%%%%%%%%%%%%%%%%%%%%%%%%%%%%%%%%%%%%%%%%%%%%%%%%%%%%%
\section{L-GATr details}
\label{gatr_details}
We summarize the details of the L-GATr trainings for the tasks presented in Section~\ref{comparison}. The hyperparameter setups are summarized in Table~\ref{tab:gatr_build}.

\begin{table}[H]
%\fontsize{7}{7}\selectfont
    \centering
    \setlength{\tabcolsep}{4pt}
    \resizebox{\columnwidth}{!}{
    \begin{tabular}{l cccc}
        \toprule
        Hyperparameter & Unfolding (L-GATr) & Unfolding (L-GATr-slim) & HERA Classification & Anomaly detection \\[1mm]
        \midrule
        Architecture & \makecell{32 scalar ch. \\ 16 multivector ch. \\ 12 blocks \\ 8 heads} &  \makecell{64 scalar ch. \\ 32 vector ch. \\ 12 blocks \\ 4 heads} &  \makecell{32 scalar ch. \\ 16 multivector ch. \\ 12 blocks \\ 8 heads} & \makecell{32 scalar ch. \\ 16 multivector ch. \\ 12 blocks \\ 8 heads}  \\
        Parameters & $10^6$ & $7 \times 10^5$ & $10^6$ & $1.8 \times 10^6$ \\
        \midrule        
        Optimizer & Lion~\cite{chen2023symbolic} & Lion~\cite{chen2023symbolic} & Lion~\cite{chen2023symbolic} & AdamW~\cite{loshchilov2017decoupled} \\
        Learning rate & $3 \times 10^{-5} (10^{-6})$ & $2 \times 10^{-6} (10^{-6})$ & $3 \times 10^{-5}$ & $5 \times 10^{-4}$ \\
        Weight decay & $0.1$ & $0.1$ & $0.2$ & $10^{-3}$ \\
        Batch size & 512 & 1024 & 512 & 1024 \\
Scheduler & Cosine Decay & Cosine Decay & Cosine Decay & Cosine Decay \\      
        Epochs & 20 & 20 & 20 & 60 
        \\
        \bottomrule
    \end{tabular}} 
    \caption{Hyperparameter summary for all trainings performed for Section~\ref{comparison}. We include a description for both L-GATr and L-GATr-slim networks trained for the reweighting-based unfolding task presented in Section~\ref{zjets}. We remark that wide hyperparameter changes on the unfolding and HERA classification tasks result in similar performance. Here we just indicate the hyperparameter setup that best matches \textsc{OmniLearn}. The learning rates in parentheses denote the fixed values used for the later iterations of the unfolding procedure.} 
    \label{tab:gatr_build}
\end{table}

\subsubsection*{Reweighting-based unfolding for pp collisions}
\label{hyperparameters_zjets}

We discuss here the training setup that we use for the L-GATr and L-GATr-slim networks in Section~\ref{zjets}. The L-GATr architecture with $10^6$ parameters consists of 16 multivector channels, 32 scalar channels, 4 attention heads and 12 transformer blocks. The $2\times 10^6$ parameter network increases the number of transformer blocks to 24. Variations on these network shapes do not have a visible effect on both reweighting and unfolding performance. All networks are trained for 20 epochs and validated every epoch. No improvement is observed from increasing the training time to 40 epochs. 

Concerning the hyperparameter setup, we perform the following sweep of training parameter values:

\begin{itemize}
    \item Learning rate: $\{10^5, \mathbf{3\times 10^5}, 10^4\}$ 
    \item Batch size: $\{\mathbf{512}, 1024, 2048\}$ 
    \item Scheduler: \{No scheduler, Cosine Annealing, \textbf{Cosine Decay}\} 
    \item Weight decay: $\{0, 0.05, \mathbf{0.1}\}$
\end{itemize}

We highlight the hyperparameter values that are used on the OmniLearn setup in Ref.~\cite{Mikuni:2024qsr}. The Cosine Decay setup matches exactly the OmniLearn configuration. We experiment with changes on the learning rate and scheduler just on the first unfolding iteration. We fix the learning rate to $10^{-6}$ on the later iterations, following once again the OmniLearn prescription. We run all of our trainings with the Lion optimizer~\cite{chen2023symbolic}. We perform 6 independent trainings for each combination of these values and display the top performer on each task in Tables~\ref{tab:triangle-metrics-reco} and~\ref{tab:triangle-metrics-hard} and Figs.~\ref{fig:reweighted_results} and~\ref{fig:unfolded_results}. 

As for the L-GATr-slim network, we apply the prescription outlined in Ref.~\cite{Petitjean:2025zjf} to adapt the L-GATr setup with $10^6$ parameters for the simplified architecture, resulting in a build with 32 vector channels, 64 scalar channels, 8 attention heads and 12 network blocks, amounting to a total of $7 \times 10^5$ parameters. We observe performance instability upon changing the network shape or size.  

Concerning hyper-parameter optimization, the results from L-GATr-slim are much more predictable under different training setups. More specifically, changing the scheduler and weight decay from the OmniLearn reference results in a clear performance degradation, whereas we observe that the optimal learning rate is $2\times 10^{-6}$. By contrast, we observe a very weak dependence on the batch size, prompting us to run all our trainings with a fixed value of 1024. We also find that performance improves significantly when including scalar inputs. We run three repeated trainings with the optimal setup and display the best results on Tables~\ref{tab:triangle-metrics-reco} and~\ref{tab:triangle-metrics-hard}. 

\subsubsection*{Likelihood ratio estimation for $ep$ collisions}
\label{hyperparameters_h1}

We discuss the scan over the hyper-parameters of the L-GATr model performed for the training setup in Section~\ref{h1}. The $10^6$ parameter network consists of 16 multivector channels, 32 scalar channels, 8 attention heads and 12 transformer blocks. The $2\times 10^6$ parameter network changed the number of multivector channels to 20, scalar channels to 40, 4 attention heads and 15 transformer blocks. All of our networks are trained for 20 epochs, and no improvement is observed from increasing the training time to 40 epochs. 

We also perform a scan over the following hyperparameters:

\begin{itemize}
    \item Learning rate: $\{10^{-5}, \mathbf{3\times 10^{-5}}, 10^{-4}\}$ 
    \item Batch size: ${128, 256, \mathbf{512}}$ 
    \item Weight decay: $\{0, 0.05, \mathbf{0.2}\}$
\end{itemize}
We run all of our trainings with the Lion optimizer. We perform 6 independent trainings for each hyper-parameter, while keeping the other hyper-parameters fixed to the \textsc{OmniLearn} values, and display the top performer in Table~\ref{tab:classifier_performance}.

\subsubsection*{Weakly supervised anomaly detection}
\label{hyperparameters_ad}

The L-GATr network used in Section~\ref{ad} consists of 16 multivector channels, 32 scalar channels, 8 attention heads and 12 transformer blocks, amounting to a total of about $1.8 \times 10^6$ parameters. Training a smaller network with about $1.4 \times 10^5$ parameters produces significantly worse results.

Unlike the rest of our studies, we observe a weak dependence on the rest of the hyperparameters for this task, so there is no need to run a scan to maximize performance. We train the network with the following values:

\begin{itemize}
    \item Learning rate: $5 \times 10^{-4}$
    \item Batch size: 1024
    \item Weight decay: $10^{-3}$
\end{itemize}
The training lasts for 60 epochs or until the validation loss fails to decrease for 10 epochs, and we use the AdamW optimizer.

\bibliographystyle{JHEP} % or try abbrvnat or unsrtnat
\bibliography{tilman, HEPML, other, EEC_ref} % refers to example.bib

@techreport{pythiatune,
      title         = "{{ATLAS Pythia 8 Tunes to 7 TeV Data}}",
      institution   = "CERN",
      reportNumber  = "ATL-PHYS-PUB-2014-021",
      address       = "Geneva",
      year          = "2014",
      url           = "https://cds.cern.ch/record/1966419",
      note          = "All figures including auxiliary figures are available at
                       https://atlas.web.cern.ch/Atlas/GROUPS/PHYSICS\\/PUBNOTES/ATL-PHYS-PUB-2014-021",
}

@article{omnifold,
	doi = {10.1103/physrevlett.124.182001},
  
	url = {https://doi.org/10.1103%2Fphysrevlett.124.182001},
  
	year = 2020,
	month = {May},
  
	publisher = {American Physical Society ({APS})},
  
	volume = {124},
  
	number = {18},
  
	author = {Anders Andreassen and Patrick T. Komiske and Eric M. Metodiev and Benjamin Nachman and Jesse Thaler},
  
	title = {{{OmniFold}: A Method to Simultaneously Unfold All Observables}},
  
	journal = {Physical Review Letters}
}

@article{Chappell:2022yxd,
    author = "Chappell, Andrew and Whitehead, Leigh H.",
    title = "{Application of transfer learning to neutrino interaction classification}",
    eprint = "2207.03139",
    archivePrefix = "arXiv",
    primaryClass = "hep-ex",
    doi = "10.1140/epjc/s10052-022-11066-6",
    journal = "Eur. Phys. J. C",
    volume = "82",
    number = "12",
    pages = "1099",
    year = "2022"
}

@misc{LHCOlympics,
   author       = "Gregor Kasieczka and Benjamin Nachman and David Shih",
   year         = "2019",
   title       = "{Official Datasets for LHC Olympics 2020 Anomaly Detection Challenge (Version v6) [Data set].}",
   publisher    = {Zenodo},
   note = "https://doi.org/10.5281/zenodo.4536624",
}

@article{Buhmann:2023acn,
    author = "Buhmann, Erik and Ewen, Cedric and Kasieczka, Gregor and Mikuni, Vinicius and Nachman, Benjamin and Shih, David",
    title = "{Full Phase Space Resonant Anomaly Detection}",
    eprint = "2310.06897",
    archivePrefix = "arXiv",
    primaryClass = "hep-ph",
    month = "10",
    year = "2023"
}

@inproceedings{Bogatskiy:2022hub,
    author = "Bogatskiy, Alexander and others",
    title = "{Symmetry Group Equivariant Architectures for Physics}",
    booktitle = "{2022 Snowmass Summer Study}",
    eprint = "2203.06153",
    archivePrefix = "arXiv",
    primaryClass = "cs.LG",
    month = "3",
    year = "2022"
}

@article{Arratia:2021tsq,
    author = "Arratia, Miguel and Britzger, Daniel and Long, Owen and Nachman, Benjamin",
    title = "{Reconstructing the Kinematics of Deep Inelastic Scattering with Deep Learning}",
    eprint = "2110.05505",
    archivePrefix = "arXiv",
    primaryClass = "hep-ex",
    reportNumber = "MPP-2021-174",
    month = "10",
    year = "2021"
}

@article{Hallin:2021wme,
    author = "Hallin, Anna and Isaacson, Joshua and Kasieczka, Gregor and Krause, Claudius and Nachman, Benjamin and Quadfasel, Tobias and Schlaffer, Matthias and Shih, David and Sommerhalder, Manuel",
    title = "{Classifying Anomalies THrough Outer Density Estimation (CATHODE)}",
    eprint = "2109.00546",
    archivePrefix = "arXiv",
    primaryClass = "hep-ph",
    reportNumber = "EFI-20-5, FERMILAB-PUB-21-389-T",
    month = "9",
    year = "2021"
}

@article{Aarrestad:2021oeb,
    author = "Aarrestad, T. and others",
    title = "{The Dark Machines Anomaly Score Challenge: Benchmark Data and Model Independent Event Classification for the Large Hadron Collider}",
    eprint = "2105.14027",
    archivePrefix = "arXiv",
    primaryClass = "hep-ph",
    month = "5",
    year = "2021"
}

@article{Andreassen:2021zzk,
    author = "Andreassen, Anders and Komiske, Patrick T. and Metodiev, Eric M. and Nachman, Benjamin and Suresh, Adi and Thaler, Jesse",
    title = "{Scaffolding Simulations with Deep Learning for High-dimensional Deconvolution}",
    eprint = "2105.04448",
    archivePrefix = "arXiv",
    primaryClass = "stat.ML",
    month = "5",
    year = "2021"
}

@article{Feickert:2021ajf,
    author = "Feickert, Matthew and Nachman, Benjamin",
    title ="{A Living Review of Machine Learning for Particle Physics}",
    eprint = "2102.02770",
    archivePrefix = "arXiv",
    primaryClass = "hep-ph",
    month = "2",
    year = "2021"
}

@article{Kasieczka:2021xcg,
    author = "Kasieczka, Gregor and others",
    title = "{The LHC Olympics 2020: A Community Challenge for Anomaly Detection in High Energy Physics}",
    eprint = "2101.08320",
    archivePrefix = "arXiv",
    primaryClass = "hep-ph",
    month = "1",
    year = "2021"
}

@article{Nachman:2020lpy,
    author = "Nachman, Benjamin and Shih, David",
    title = "{Anomaly Detection with Density Estimation}",
    eprint = "2001.04990",
    archivePrefix = "arXiv",
    primaryClass = "hep-ph",
    doi = "10.1103/PhysRevD.101.075042",
    journal = "Phys. Rev. D",
    volume = "101",
    pages = "075042",
    year = "2020"
}

@article{Radovic:2018dip,
    author = "Radovic, Alexander and Williams, Mike and Rousseau, David and Kagan, Michael and Bonacorsi, Daniele and Himmel, Alexander and Aurisano, Adam and Terao, Kazuhiro and Wongjirad, Taritree",
    doi = "10.1038/s41586-018-0361-2",
    journal = "Nature",
    number = "7716",
    pages = "41--48",
    reportNumber = "FERMILAB-PUB-18-436-ND",
    title = "{Machine learning at the energy and intensity frontiers of particle physics}",
    volume = "560",
    year = "2018"
}

@article{Andreassen:2019cjw,
    author = "Andreassen, Anders and Komiske, Patrick T. and Metodiev, Eric M. and Nachman, Benjamin and Thaler, Jesse",
    title = "{OmniFold: A Method to Simultaneously Unfold All Observables}",
    eprint = "1911.09107",
    archivePrefix = "arXiv",
    primaryClass = "hep-ph",
    reportNumber = "MIT-CTP 5155",
    doi = "10.1103/PhysRevLett.124.182001",
    journal = "Phys. Rev. Lett.",
    volume = "124",
    number = "18",
    pages = "182001",
    year = "2020"
}

@article{Metodiev:2017vrx,
    author = "Metodiev, Eric M. and Nachman, Benjamin and Thaler, Jesse",
    title = "{Classification without labels: Learning from mixed samples in high energy physics}",
    eprint = "1708.02949",
    archivePrefix = "arXiv",
    primaryClass = "hep-ph",
    reportNumber = "MIT--CTP-4922",
    doi = "10.1007/JHEP10(2017)174",
    journal = "JHEP",
    volume = "10",
    pages = "174",
    year = "2017"
}

@article{Gambhir:2025afb,
    author = "Gambhir, Rikab and Mastandrea, Radha and Nachman, Benjamin and Thaler, Jesse",
    title = "{Isolating Unisolated Upsilons with Anomaly Detection in CMS Open Data}",
    eprint = "2502.14036",
    archivePrefix = "arXiv",
    primaryClass = "hep-ph",
    reportNumber = "MIT-CTP 5843",
    doi = "10.1103/vvv3-5kkl",
    journal = "Phys. Rev. Lett.",
    volume = "135",
    number = "2",
    pages = "021902",
    year = "2025"
}

@misc{omnifold_zenodo,
  author       = {Anders Andreassen and Patrick T. Komiske and Eric M. Metodiev and Benjamin Nachman and Jesse Thaler},
  title        = {Pythia/Herwig + Delphes Jet Datasets for OmniFold Unfolding},
  year         = {2019},
  publisher    = {Zenodo},
  doi          = {10.5281/zenodo.3548091},
  url          = {https://zenodo.org/record/3548091}
}

@article{Cowan:2002in,
    author = "Cowan, G.",
    editor = "Whalley, M. R. and Lyons, L.",
    title = "{A survey of unfolding methods for particle physics}",
    journal = "Conf. Proc. C",
    volume = "0203181",
    pages = "248--257",
    year = "2002"
}

@inproceedings{Blobel:2011fih,
    author = "Blobel, Volker",
    title = "{Unfolding Methods in Particle Physics}",
    booktitle = "{PHYSTAT 2011}",
    doi = "10.5170/CERN-2011-006.240",
    publisher = "CERN",
    address = "Geneva",
    pages = "240--251",
    year = "2011"
}

@book{Behnke:2013pga,
    editor = {Behnke, Olaf and Kr{\"o}ninger, Kevin and Sch{\"o}rner-Sadenius, Thomas and Schott, Gregory},
    title = "{Data analysis in high energy physics}: {A practical guide to statistical methods}",
    isbn = "978-3-527-41058-3, 978-3-527-65344-7, 978-3-527-65343-0",
    publisher = "Wiley-VCH",
    address = "Weinheim, Germany",
    year = "2013"
}

@article{Brenner:2019lmf,
    author = "Brenner, Lydia and Balasubramanian, Rahul and Burgard, Carsten and Verkerke, Wouter and Cowan, Glen and Verschuuren, Pim and Croft, Vincent",
    title = "{Comparison of unfolding methods using RooFitUnfold}",
    eprint = "1910.14654",
    archivePrefix = "arXiv",
    primaryClass = "physics.data-an",
    doi = "10.1142/S0217751X20501456",
    journal = "Int. J. Mod. Phys. A",
    volume = "35",
    number = "24",
    pages = "2050145",
    year = "2020"
}

@article{DAGOSTINI1995487,
title = {A multidimensional unfolding method based on Bayes' theorem},
journal = {Nuclear Instruments and Methods in Physics Research Section A: Accelerators, Spectrometers, Detectors and Associated Equipment},
volume = {362},
number = {2},
pages = {487-498},
year = {1995},
issn = {0168-9002},
doi = {https://doi.org/10.1016/0168-9002(95)00274-X},
url = {https://www.sciencedirect.com/science/article/pii/016890029500274X},
author = {G. D'Agostini},
}

@article{Qu:2022mxj,
    author = "Qu, Huilin and Li, Congqiao and Qian, Sitian",
    title = "{Particle Transformer for Jet Tagging}",
    eprint = "2202.03772",
    archivePrefix = "arXiv",
    primaryClass = "hep-ph",
    month = "2",
    year = "2022"
}

@article{Mikuni:2024qsr,
    author = "Mikuni, Vinicius and Nachman, Benjamin",
    title = "{Solving key challenges in collider physics with foundation models}",
    eprint = "2404.16091",
    archivePrefix = "arXiv",
    primaryClass = "hep-ph",
    doi = "10.1103/PhysRevD.111.L051504",
    journal = "Phys. Rev. D",
    volume = "111",
    number = "5",
    pages = "L051504",
    year = "2025"
}

@article{loshchilov2017decoupled,
  title={Decoupled weight decay regularization},
  author={Loshchilov, Ilya and Hutter, Frank},
  journal={arXiv preprint arXiv:1711.05101},
  year={2017}
}

@article{Batson:2023ohn,
    author = "Batson, Joshua and Kahn, Yonatan Frederick",
    title = "{Scaling laws in jet classification}",
    eprint = "2312.02264",
    archivePrefix = "arXiv",
    primaryClass = "hep-ph",
    doi = "10.21468/SciPostPhysCore.8.1.034",
    journal = "SciPost Phys. Core",
    volume = "8",
    pages = "034",
    year = "2025"
}

@inproceedings{Bogatskiy:2023fug,
    author = "Bogatskiy, Alexander and Hoffman, Timothy and Offermann, Jan T.",
    title = "{19 Parameters Is All You Need: Tiny Neural Networks for Particle Physics}",
    booktitle = "{37th Conference on Neural Information Processing Systems}",
    eprint = "2310.16121",
    archivePrefix = "arXiv",
    primaryClass = "hep-ph",
    month = "10",
    year = "2023"
}

@article{Vigl:2026ppx,
    author = "Vigl, Matthias and Hartman, Nicole and Kagan, Michael and Heinrich, Lukas",
    title = "{Neural Scaling Laws for Boosted Jet Tagging}",
    eprint = "2602.15781",
    archivePrefix = "arXiv",
    primaryClass = "hep-ex",
    month = "2",
    year = "2026"
}

@misc{brehmer2023geometricalgebratransformer,
      title={Geometric Algebra Transformer}, 
      author={Johann Brehmer and Pim de Haan and Sönke Behrends and Taco Cohen},
      year={2023},
      eprint={2305.18415},
      archivePrefix={arXiv},
      primaryClass={cs.LG},
      url={https://arxiv.org/abs/2305.18415}, 
}

@article{Belis:2023mqs,
    author = "Belis, Vasilis and Odagiu, Patrick and Aarrestad, Thea Klaeboe",
    title = "{Machine learning for anomaly detection in particle physics}",
    eprint = "2312.14190",
    archivePrefix = "arXiv",
    primaryClass = "physics.data-an",
    doi = "10.1016/j.revip.2024.100091",
    journal = "Rev. Phys.",
    volume = "12",
    pages = "100091",
    year = "2024"
}

@article{Hebbar:2025adf,
    author = "Hebbar, Pradyun and Madula, Thandikire and Mikuni, Vinicius and Nachman, Benjamin and Outmezguine, Nadav and Savoray, Inbar",
    title = "{SEAL - A Symmetry EncourAging Loss for High Energy Physics}",
    eprint = "2511.01982",
    archivePrefix = "arXiv",
    primaryClass = "hep-ph",
    month = "11",
    year = "2025"
}

@article{Nabat:2024nce,
    author = "Nabat, Seth and Ghosh, Aishik and Witkowski, Edmund and Kasieczka, Gregor and Whiteson, Daniel",
    title = "{Learning broken symmetries with approximate invariance}",
    eprint = "2412.18773",
    archivePrefix = "arXiv",
    primaryClass = "hep-ph",
    doi = "10.1103/PhysRevD.111.072002",
    journal = "Phys. Rev. D",
    volume = "111",
    number = "7",
    pages = "072002",
    year = "2025"
}

@article{Dillon:2022tmm,
    author = "Dillon, Barry M. and Mastandrea, Radha and Nachman, Benjamin",
    title = "{Self-supervised anomaly detection for new physics}",
    eprint = "2205.10380",
    archivePrefix = "arXiv",
    primaryClass = "hep-ph",
    journal = "Phys. Rev. D",
    volume = "106",
    pages = "056005",
    year = "2022"
}

@article{Harris:2024sra,
    author = "Harris, Philip and Kagan, Michael and Krupa, Jeffrey and Maier, Blaize and Woodward, Nhan",
    title = "{Foundation models for particle physics event reconstruction}",
    eprint = "2403.07066",
    archivePrefix = "arXiv",
    primaryClass = "hep-ph",
    year = "2024"
}

@article{Kuchera:2018msh,
    author = "Kuchera, M. P. and Ramanujan, R. and Taylor, J. Z. and Strauss, R. R. and Bazin, D. and Bradt, J. and Chen, R.",
    title = "{Machine learning methods for track classification in the AT-TPC}",
    eprint = "1810.10350",
    archivePrefix = "arXiv",
    primaryClass = "cs.CV",
    journal = "Nucl. Instrum. Meth. A",
    volume = "940",
    pages = "156",
    year = "2019"
}

@article{Dreyer:2022yom,
    author = "Dreyer, Fr\'ed\'eric A. and Grabarczyk, Reinhard and Monni, Pier Francesco",
    title = "{Leveraging universality of jet taggers through transfer learning}",
    eprint = "2203.06210",
    archivePrefix = "arXiv",
    primaryClass = "hep-ph",
    journal = "Eur. Phys. J. C",
    volume = "82",
    pages = "564",
    year = "2022"
}

@article{Beauchesne:2023xhj,
    author = "Beauchesne, Hugues and Chen, Zhong-En and Chiang, Cheng-Wei",
    title = "{Pre-training strategies using contrastive learning and corpus transfer for jet physics}",
    eprint = "2312.06152",
    archivePrefix = "arXiv",
    primaryClass = "hep-ph",
    year = "2023"
}

@article{Birk:2024knn,
    author = "Birk, Joschka and Hallin, Anna and Kasieczka, Gregor",
    title = "{OmniJet-$\alpha$: The first cross-task foundation model for particle physics}",
    eprint = "2403.05618",
    archivePrefix = "arXiv",
    primaryClass = "hep-ph",
    year = "2024"
}

@article{Hallin:2025wme,
    author = "Hallin, Anna",
    title = "{OmniJet-$\alpha$-2: A cross-task foundation model update}",
    eprint = "2509.21434",
    archivePrefix = "arXiv",
    primaryClass = "hep-ph",
    year = "2025"
}

@article{Golling:2024abg,
    author = "Golling, Tobias and Heinrich, Lukas and Kagan, Michael and Klein, Samuel and Leigh, Matthew and Osadchy, Margarita and Raine, John A.",
    title = "{Masked particle modeling on sets: Towards self-supervised high energy physics foundation models}",
    eprint = "2401.13537",
    archivePrefix = "arXiv",
    primaryClass = "hep-ph",
    journal = "Mach. Learn. Sci. Tech.",
    volume = "5",
    pages = "035074",
    year = "2024"
}

@article{Leigh:2024ked,
    author = "Leigh, Matthew and Klein, Samuel and Charton, Fran\c{c}ois and Golling, Tobias and Heinrich, Lukas and Kagan, Michael and Ochoa, In\^{e}s and Osadchy, Margarita",
    title = "{Scaling masked particle modeling on sets}",
    eprint = "2409.12589",
    archivePrefix = "arXiv",
    primaryClass = "hep-ph",
    year = "2024"
}

@article{Bardhan:2025dpn,
    author = "Bardhan, Jyotirmoy and Agrawal, Rajat and Tilak, Ashwin and Neeraj, C. and Mitra, Srikrishna",
    title = "{A self-supervised foundation model for jet physics}",
    eprint = "2502.03933",
    archivePrefix = "arXiv",
    primaryClass = "cs.LG",
    year = "2025"
}

@article{Mikuni:2025wjk,
    author = "Mikuni, Vinicius and Nachman, Benjamin",
    title = "{A Method to Simultaneously Facilitate All Jet Physics Tasks}",
    eprint = "2502.14652",
    archivePrefix = "arXiv",
    primaryClass = "hep-ph",
    journal = "Phys. Rev. D",
    volume = "111",
    pages = "054015",
    year = "2025"
}

@article{Bhimji:2025tpq,
    author = "Bhimji, Wahid and Harris, Chris and Mikuni, Vinicius and Nachman, Benjamin",
    title = "{OmniLearned: A Foundation Model Framework for All Tasks Involving Jet Physics}",
    eprint = "2510.24066",
    archivePrefix = "arXiv",
    primaryClass = "hep-ph",
    year = "2025"
}

@article{Mikuni:2025ezi,
    author = "Mikuni, Vinicius and Elsharkawy, Ibrahim and Nachman, Benjamin",
    title = "{OmniCosmos: Transferring Particle Physics Knowledge Across the Cosmos}",
    eprint = "2512.24422",
    archivePrefix = "arXiv",
    primaryClass = "astro-ph.CO",
    year = "2025"
}

@article{Elsharkawy:2026bnm,
    author = "Elsharkawy, Ibrahim and Mikuni, Vinicius and Bhimji, Wahid and Nachman, Benjamin",
    title = "{OmniMol: Transferring Particle Physics Knowledge to Molecular Dynamics with Point-Edge Transformers}",
    eprint = "2601.10791",
    archivePrefix = "arXiv",
    primaryClass = "physics.chem-ph",
    year = "2026"
}

@article{Karagiorgi:2021ngt,
    author = "Karagiorgi, Georgia and Kasieczka, Gregor and Kravitz, Scott and Nachman, Benjamin and Shih, David",
    title = "{Machine Learning in the Search for New Fundamental Physics}",
    eprint = "2112.03769",
    archivePrefix = "arXiv",
    primaryClass = "hep-ph",
    month = "12",
    year = "2021"
}

@article{Gong:2022lye,
    author = "Gong, Shiqi and Meng, Qi and Zhang, Jue and Qu, Huilin and Li, Congqiao and Qian, Sitian and Du, Weitao and Ma, Zhi-Ming and Liu, Tie-Yan",
    title = "{An efficient Lorentz equivariant graph neural network for jet tagging}",
    journal = "JHEP",
    volume = "07",
    pages = "030",
    year = "2022",
    doi = "10.1007/JHEP07(2022)030",
    eprint = "2201.08187",
    archivePrefix = "arXiv",
    primaryClass = "hep-ph"
}

@article{Qiu:2022xvr,
    author = "Qiu, Shikai and Han, Shuo and Ju, Xiangyang and Nachman, Benjamin and Wang, Haichen",
    title = "{Holistic approach to predicting top quark kinematic properties with the covariant particle transformer}",
    journal = "Phys. Rev. D",
    volume = "107",
    pages = "114029",
    year = "2023",
    doi = "10.1103/PhysRevD.107.114029",
    eprint = "2203.05687",
    archivePrefix = "arXiv",
    primaryClass = "hep-ph"
}

@article{Bogatskiy:2023nnw,
    author = "Bogatskiy, Alexander and Hoffman, Timothy and Miller, David W. and Offermann, Jan T. and Liu, Xiaoyang",
    title = "{Explainable equivariant neural networks for particle physics: PELICAN}",
    journal = "JHEP",
    volume = "03",
    pages = "113",
    year = "2024",
    doi = "10.1007/JHEP03(2024)113",
    eprint = "2307.16506",
    archivePrefix = "arXiv",
    primaryClass = "hep-ph"
}

@inproceedings{Batatia:2023nqg,
    author = "Batatia, Ilyes and Geiger, Mario and Munoz, Jose and Smidt, Tess and Silberman, Lior and Ortner, Christoph",
    title = "{A General Framework for Equivariant Neural Networks on Reductive Groups}",
    booktitle = "{Advances in Neural Information Processing Systems 36}",
    pages = "55260",
    year = "2023"
}

@inproceedings{Bogatskiy:2020tje,
    author = "Bogatskiy, Alexander and Anderson, Brandon and Offermann, Jan T. and Roussi, Marwah and Miller, David W. and Kondor, Risi",
    title = "{Lorentz Group Equivariant Neural Network for Particle Physics}",
    booktitle = "{Proceedings of the 37th International Conference on Machine Learning}",
    series = "{Proceedings of Machine Learning Research}",
    volume = "119",
    pages = "992--1002",
    year = "2020",
    eprint = "2006.04780",
    archivePrefix = "arXiv",
    primaryClass = "hep-ph"
}

@article{Bogatskiy:2022hrp,
    author = "Bogatskiy, Alexander and Hoffman, Timothy and Miller, David W. and Offermann, Jan T.",
    title = "{PELICAN: Permutation Equivariant and Lorentz Invariant or Covariant Aggregator Network for Particle Physics}",
    year = "2022",
    eprint = "2211.00454",
    archivePrefix = "arXiv",
    primaryClass = "hep-ph"
}

@article{Li:2022xfc,
    author = "Li, Congqiao and Qu, Huilin and Qian, Sitian and Meng, Qi and Gong, Shiqi and Zhang, Jue and Liu, Tie-Yan and Li, Qiang",
    title = "{Does Lorentz-symmetric design boost network performance in jet physics?}",
    journal = "Phys. Rev. D",
    volume = "109",
    pages = "056003",
    year = "2024",
    doi = "10.1103/PhysRevD.109.056003",
    eprint = "2208.07814",
    archivePrefix = "arXiv",
    primaryClass = "hep-ph"
}

@article{Elhag:2024oer,
    author = "Elhag, Ahmed A. and Rusch, T. Konstantin and Di Giovanni, Francesco and Bronstein, Michael",
    title = "{Relaxed Equivariance via Multitask Learning}",
    year = "2024",
    eprint = "2410.17878",
    archivePrefix = "arXiv",
    primaryClass = "cs.LG"
}

@inproceedings{AkhoundSadegh:2023xlp,
    author    = "Akhound-Sadegh, Tara and Perreault-Levasseur, Laurence 
                 and Brandstetter, Johannes and Welling, Max and Ravanbakhsh, Siamak",
    title     = "{Lie Point Symmetry and Physics Informed Networks}",
    booktitle = "{Advances in Neural Information Processing Systems}",
    volume    = "36",
    pages     = "42468--42481",
    year      = "2023",
    eprint    = "2311.04293",
    archivePrefix = "arXiv"
}

@article{Hao:2022zns,
    author = "Hao, Zichun and Kansal, Raghav and Duarte, Javier and Chernyavskaya, Nadezda",
    title = "{Lorentz group equivariant autoencoders}",
    journal = "Eur. Phys. J. C",
    volume = "83",
    pages = "485",
    year = "2023",
    doi = "10.1140/epjc/s10052-023-11633-5",
    eprint = "2212.07347",
    archivePrefix = "arXiv",
    primaryClass = "hep-ph"
}

@inproceedings{Villar:2021hph,
    author = "Villar, Soledad and Hogg, David W. and Storey-Fisher, Kate and Yao, Weichi and Blum-Smith, Ben",
    title = "{Scalars are universal: Equivariant machine learning, structured like classical physics}",
    booktitle = "{Advances in Neural Information Processing Systems 34}",
    year = "2021",
    eprint = "2106.06610",
    archivePrefix = "arXiv",
    primaryClass = "cs.LG"
}

@article{
doi:10.1073/pnas.1912789117,
author = {Kyle Cranmer  and Johann Brehmer  and Gilles Louppe },
title = {The frontier of simulation-based inference},
journal = {Proceedings of the National Academy of Sciences},
volume = {117},
number = {48},
pages = {30055-30062},
year = {2020},
doi = {10.1073/pnas.1912789117},
URL = {https://www.pnas.org/doi/abs/10.1073/pnas.1912789117},
eprint = {https://www.pnas.org/doi/pdf/10.1073/pnas.1912789117}}

@article{Canelli:2025ybb,
    author = "Canelli, Florencia and others",
    title = "{A practical guide to unbinned unfolding}",
    eprint = "2507.09582",
    archivePrefix = "arXiv",
    primaryClass = "hep-ph",
    doi = "10.1140/epjc/s10052-025-15265-9",
    journal = "Eur. Phys. J. C",
    volume = "86",
    number = "2",
    pages = "106",
    year = "2026"
}

@article{Long:2023mrj,
    author = "Long, Owen and Nachman, Benjamin",
    title = "{Designing observables for measurements with deep learning}",
    eprint = "2310.08717",
    archivePrefix = "arXiv",
    primaryClass = "physics.data-an",
    doi = "10.1140/epjc/s10052-024-13135-4",
    journal = "Eur. Phys. J. C",
    volume = "84",
    number = "8",
    pages = "776",
    year = "2024"
}

@mastersthesis{energyflowthesis,
    author = "Peez, Matti",
    title = "{Search for deviations from the standard model in high transverse energy processes at the electron proton collider HERA}",
    reportNumber = "DESY-THESIS-2003-023, LYCEN-2003-81, CPPM-T-2003-04",
    doi = "10.3204/DESY-THESIS-2003-023",
    type = "Other thesis",
    month = "10",
    year = "2003"
}

@mastersthesis{energyflowthesis2,
    author = "Hellwig, Susanne",
    title = "{Untersuchung der $D^* - \pi_{slow}$ Double Tagging Methode in Charmanalysen}",
    school = "Hamburg U.",
    year = "2004"
}

@mastersthesis{energyflowthesis3,
    author = "Portheault, Benjamin",
    title = "{First measurement of charged and neutral current cross sections with the polarized positron beam at HERA II and QCD-electroweak analyses}",
    reportNumber = "LAL-05-05",
    type = "Other thesis",
    month = "3",
    year = "2005"
}

@article{Arratia:2022wny,
    author = "Arratia, Miguel and Britzger, Daniel and Long, Owen and Nachman, Benjamin",
    title = "{Optimizing observables with machine learning for better unfolding}",
    eprint = "2203.16722",
    archivePrefix = "arXiv",
    primaryClass = "hep-ex",
    reportNumber = "MPP-2022-35",
    doi = "10.1088/1748-0221/17/07/P07009",
    journal = "JINST",
    volume = "17",
    number = "07",
    pages = "P07009",
    year = "2022"
}

@article{Brun:1987ma,
    author = "Brun, R. and Bruyant, F. and Maire, M. and McPherson, A. C. and Zanarini, P.",
    title = "{GEANT3}",
    journal = "CERN-DD-EE-84-01",
    month = "9",
    year = "1987"
}

@article{Charchula:1994kf,
    author = "Charchula, K. and Schuler, G. A. and Spiesberger, H.",
    title = "{Combined QED and QCD radiative effects in deep inelastic lepton - proton scattering: The Monte Carlo generator DJANGO6}",
    reportNumber = "CERN-TH-7133-94",
    doi = "10.1016/0010-4655(94)90086-8",
    journal = "Comput. Phys. Commun.",
    volume = "81",
    pages = "381--402",
    year = "1994"
}

@article{Jung:1993gf,
    author = "Jung, Hannes",
    title = "{Hard diffractive scattering in high-energy e p collisions and the Monte Carlo generator RAPGAP}",
    reportNumber = "DESY-93-182",
    doi = "10.1016/0010-4655(94)00150-Z",
    journal = "Comput. Phys. Commun.",
    volume = "86",
    pages = "147--161",
    year = "1995"
}

@article{Cacciari:2005hq,
    author = "Cacciari, Matteo and Salam, Gavin P.",
    title = "{Dispelling the $N^{3}$ myth for the $k_t$ jet-finder}",
    eprint = "hep-ph/0512210",
    archivePrefix = "arXiv",
    reportNumber = "LPTHE-05-32",
    doi = "10.1016/j.physletb.2006.08.037",
    journal = "Phys. Lett. B",
    volume = "641",
    pages = "57--61",
    year = "2006"
}

@article{Bahr:2008pv,
    author = "Bahr, M. and others",
    title = "{Herwig++ Physics and Manual}",
    eprint = "0803.0883",
    archivePrefix = "arXiv",
    primaryClass = "hep-ph",
    reportNumber = "CERN-PH-TH-2008-038, CAVENDISH-HEP-08-03, KA-TP-05-2008, DCPT-08-22, IPPP-08-11, CP3-08-05",
    doi = "10.1140/epjc/s10052-008-0798-9",
    journal = "Eur. Phys. J. C",
    volume = "58",
    pages = "639--707",
    year = "2008"
}

@article{Bellm:2015jjp,
    author = "Bellm, Johannes and others",
    title = "{Herwig 7.0/Herwig++ 3.0 release note}",
    eprint = "1512.01178",
    archivePrefix = "arXiv",
    primaryClass = "hep-ph",
    reportNumber = "CERN-PH-TH-2015-289, MAN-HEP-2015-15, IFJPAN-IV-2015-13, KA-TP-18-2015, DCPT-15-142, MCNET-15-28, IPPP-15-71, HERWIG-2015-01",
    doi = "10.1140/epjc/s10052-016-4018-8",
    journal = "Eur. Phys. J. C",
    volume = "76",
    number = "4",
    pages = "196",
    year = "2016"
}

@article{Bellm:2017bvx,
    author = "Bellm, Johannes and others",
    title = "{Herwig 7.1 Release Note}",
    eprint = "1705.06919",
    archivePrefix = "arXiv",
    primaryClass = "hep-ph",
    reportNumber = "CERN-PH-TH-2017-109, CERN-TH-2017-109, MAN-HEP-2017-08, UWTHPH-2017-10, IFJPAN-IV-2017-7, NIKHEF-2017-026, HERWIG-2017-02, KA-TP-19-2017, MCNET-17-08, IPPP-17-40",
    month = "5",
    year = "2017"
}

@article{Sjostrand:2014zea,
    author = {Sj\"ostrand, Torbj\"orn and Ask, Stefan and Christiansen, Jesper R. and Corke, Richard and Desai, Nishita and Ilten, Philip and Mrenna, Stephen and Prestel, Stefan and Rasmussen, Christine O. and Skands, Peter Z.},
    title = "{An introduction to PYTHIA 8.2}",
    eprint = "1410.3012",
    archivePrefix = "arXiv",
    primaryClass = "hep-ph",
    reportNumber = "LU-TP-14-36, MCNET-14-22, CERN-PH-TH-2014-190, FERMILAB-PUB-14-316-CD, DESY-14-178, SLAC-PUB-16122",
    doi = "10.1016/j.cpc.2015.01.024",
    journal = "Comput. Phys. Commun.",
    volume = "191",
    pages = "159--177",
    year = "2015"
}

@article{Cacciari:2011ma,
    author = "Cacciari, Matteo and Salam, Gavin P. and Soyez, Gregory",
    title = "{FastJet User Manual}",
    eprint = "1111.6097",
    archivePrefix = "arXiv",
    primaryClass = "hep-ph",
    reportNumber = "CERN-PH-TH-2011-297",
    doi = "10.1140/epjc/s10052-012-1896-2",
    journal = "Eur. Phys. J. C",
    volume = "72",
    pages = "1896",
    year = "2012"
}

@article{Cacciari:2008gp,
    author         = "Cacciari, Matteo and Salam, Gavin P. and Soyez, Gregory",
    title          = "{The anti-$k_t$ jet clustering algorithm}",
    journal        = "JHEP",
    volume         = "04",
    year           = "2008",
    pages          = "063",
doi            = "10.1088/1126-6708/2008/04/063",
eprint         = "0802.1189",
archivePrefix  = "arXiv",
primaryClass   = "hep-ph",
reportNumber   = "LPTHE-07-03",
    SLACcitation   = "%%CITATION = ARXIV:0802.1189;%%"
}

@article{Selvaggi:2014mya,
author         = "Selvaggi, Michele",
    title          = "{DELPHES 3: A modular framework for fast-simulation of
    generic collider experiments}",
    booktitle      = "{Proceedings,  15th International Workshop on Advanced
    Computing and Analysis Techniques in Physics Research
    (ACAT 2013): Beijing, China, May 16-21, 2013}",
    journal        = "J. Phys. Conf. Ser.",
    volume         = "523",
    year           = "2014",
    pages          = "012033",
    doi            = "10.1088/1742-6596/523/1/012033",
    SLACcitation   = "%%CITATION = 00462,523,012033;%%"
}

@article{Mertens:2015kba,
    author         = "Mertens, Alexandre",
    title          = "{New features in Delphes 3}",
    booktitle      = "{Proceedings, 16th International workshop on Advanced
    Computing and Analysis Techniques in physics (ACAT 14):
    Prague, Czech Republic, September 1-5, 2014}",
    journal        = "J. Phys. Conf. Ser.",
    volume         = "608",
    year           = "2015",
    pages          = "012045",
    doi            = "10.1088/1742-6596/608/1/012045",
    SLACcitation   = "%%CITATION = 00462,608,012045;%%"
}

@article{deFavereau:2013fsa,
    Archiveprefix = {arXiv},
    Author = {de Favereau, J. and Delaere, C. and Demin, P. and Giammanco, A. and Lema{\^\i}tre, V. and Mertens, A. and Selvaggi, M.},
    Collaboration = {DELPHES 3},
    Doi = {10.1007/JHEP02(2014)057},
    Eprint = {1307.6346},
    Journal = {JHEP},
    Pages = {057},
    Primaryclass = {hep-ex},
    Slaccitation = {%%CITATION = ARXIV:1307.6346;%%},
    Title = {{DELPHES 3, A modular framework for fast simulation of a generic collider experiment}},
    Volume = {02},
    Year = {2014}}

@article{Hocker:1995kb,
    author         = "H{\"o}cker
    , Andreas and Kartvelishvili, Vakhtang",
    title          = "{SVD approach to data unfolding}",
    journal        = "Nucl. Instrum. Meth. A",
volume         = "372",
year           = "1996",
pages          = "469-481",
    doi            = "10.1016/0168-9002(95)01478-0",
    eprint         = "hep-ph/9509307",
    archivePrefix  = "arXiv",
    primaryClass   = "hep-ph",
    reportNumber   = "M-C-TH-95-15, LAL-95-55",
    SLACcitation   = "%%CITATION = HEP-PH/9509307;%%"
}

@article{Schmitt:2012kp,
    author         = "Schmitt, Stefan",
    title          = "{TUnfold: an algorithm for correcting migration effects
    in high energy physics}",
    journal        = "JINST",
    volume         = "7",
    year           = "2012",
    pages          = "T10003",
    doi            = "10.1088/1748-0221/7/10/T10003",
    eprint         = "1205.6201",
    archivePrefix  = "arXiv",
    primaryClass   = "physics.data-an",
    reportNumber   = "DESY-12-129",
    SLACcitation   = "%%CITATION = ARXIV:1205.6201;%%"
}

@article{Gras:2017jty,
    author = {Gras, Philippe and H\"oche, Stefan and Kar, Deepak and Larkoski, Andrew and L\"onnblad, Leif and Pl\"atzer, Simon and Si\'odmok, Andrzej and Skands, Peter and Soyez, Gregory and Thaler, Jesse},
    title = "{Systematics of quark/gluon tagging}",
    eprint = "1704.03878",
    archivePrefix = "arXiv",
    primaryClass = "hep-ph",
    reportNumber = "MIT-CTP-4885, COEPP-MN-17-2, MCNET-17-04",
    doi = "10.1007/JHEP07(2017)091",
    journal = "JHEP",
    volume = "07",
    pages = "091",
    year = "2017"
}

@article{chen2023symbolic,
      title={Symbolic Discovery of Optimization Algorithms}, 
      author={Xiangning Chen and Chen Liang and Da Huang and Esteban Real and Kaiyuan Wang and Yao Liu and Hieu Pham and Xuanyi Dong and Thang Luong and Cho-Jui Hsieh and Yifeng Lu and Quoc V. Le},
      year={2023},
      eprint={2302.06675},
      archivePrefix={arXiv},
      primaryClass={cs.LG}
}

@ARTICLE{850703,
  author={Topsoe, F.},
  journal={IEEE Transactions on Information Theory}, 
  title={Some inequalities for information divergence and related measures of discrimination}, 
  year={2000},
  volume={46},
  number={4},
  pages={1602-1609},
  doi={10.1109/18.850703}}

@article{Bright-Thonney:2018mxq,
    author = "Bright-Thonney, Samuel and Nachman, Benjamin",
    title = "{Investigating the Topology Dependence of Quark and Gluon Jets}",
    eprint = "1810.05653",
    archivePrefix = "arXiv",
    primaryClass = "hep-ph",
    doi = "10.1007/JHEP03(2019)098",
    journal = "JHEP",
    volume = "03",
    pages = "098",
    year = "2019"
}

@article{Ore:2026qgp,
    author = "Ore, Ayodele and Plehn, Tilman",
    title = "{Unfolding without Iterations, Adversaries, or Surrogates}",
    eprint = "2602.24282",
    archivePrefix = "arXiv",
    primaryClass = "hep-ph",
    month = "2",
    year = "2026"
}

@article{Petitjean:2025zjf,
    author = {Petitjean, Antoine and Plehn, Tilman and Spinner, Jonas and K{\"o}the, Ullrich},
    title = "{Economical Jet Taggers -- Equivariant, Slim, and Quantized}",
    eprint = "2512.17011",
    archivePrefix = "arXiv",
    primaryClass = "hep-ph",
    reportNumber = "IPPP/25/93",
    month = "12",
    year = "2025"
}

@article{Petitjean:2025tgk,
    author = "Petitjean, Antoine and Butter, Anja and Greif, Kevin and Palacios Schweitzer, Sofia and Plehn, Tilman and Spinner, Jonas and Whiteson, Daniel",
    title = "{Generative Unfolding of Jets and Their Substructure}",
    eprint = "2510.19906",
    archivePrefix = "arXiv",
    primaryClass = "hep-ph",
    month = "10",
    year = "2025"
}

@article{Favaro:2025pgz,
    author = "Favaro, Luigi and Gerhartz, Gerrit and Hamprecht, Fred A. and Lippmann, Peter and Pitz, Sebastian and Plehn, Tilman and Qu, Huilin and Spinner, Jonas",
    title = "{Lorentz-Equivariance without Limitations}",
    eprint = "2508.14898",
    archivePrefix = "arXiv",
    primaryClass = "hep-ph",
    month = "8",
    year = "2025"
}

@article{Bahl:2024gyt,
    author = "Bahl, Henning and Elmer, Nina and Favaro, Luigi and Haussmann, Manuel and Plehn, Tilman and Winterhalder, Ramon",
    title = "{Accurate surrogate amplitudes with calibrated uncertainties}",
    eprint = "2412.12069",
    archivePrefix = "arXiv",
    primaryClass = "hep-ph",
    doi = "10.21468/SciPostPhysCore.8.4.073",
    journal = "SciPost Phys. Core",
    volume = "8",
    pages = "073",
    year = "2025"
}

@article{Brehmer:2024yqw,
    author = "Brehmer, Johann and Bres{\'o}, V{\'\i}ctor and de Haan, Pim and Plehn, Tilman and Qu, Huilin and Spinner, Jonas and Thaler, Jesse",
    title = "{A Lorentz-equivariant transformer for all of the LHC}",
    eprint = "2411.00446",
    archivePrefix = "arXiv",
    primaryClass = "hep-ph",
    reportNumber = "MIT-CTP/5802",
    doi = "10.21468/SciPostPhys.19.4.108",
    journal = "SciPost Phys.",
    volume = "19",
    number = "4",
    pages = "108",
    year = "2025"
}

@article{Huetsch:2024quz,
    author = "Huetsch, Nathan and others",
    title = "{The landscape of unfolding with machine learning}",
    eprint = "2404.18807",
    archivePrefix = "arXiv",
    primaryClass = "hep-ph",
    doi = "10.21468/SciPostPhys.18.2.070",
    journal = "SciPost Phys.",
    volume = "18",
    number = "2",
    pages = "070",
    year = "2025"
}

@article{Dillon:2023zac,
    author = "Dillon, Barry M. and Favaro, Luigi and Feiden, Friedrich and Modak, Tanmoy and Plehn, Tilman",
    title = "{Anomalies, representations, and self-supervision}",
    eprint = "2301.04660",
    archivePrefix = "arXiv",
    primaryClass = "hep-ph",
    doi = "10.21468/SciPostPhysCore.7.3.056",
    journal = "SciPost Phys. Core",
    volume = "7",
    pages = "056",
    year = "2024"
}

@article{Plehn:2022ftl,
    author = "Plehn, Tilman and Butter, Anja and Dillon, Barry and Heimel, Theo and Krause, Claudius and Winterhalder, Ramon",
    title = "{Modern Machine Learning for LHC Physicists}",
    eprint = "2211.01421",
    archivePrefix = "arXiv",
    primaryClass = "hep-ph",
    month = "11",
    year = "2022"
}

@article{Dillon:2021gag,
    author = "Dillon, Barry M. and Kasieczka, Gregor and Olischlager, Hans and Plehn, Tilman and Sorrenson, Peter and Vogel, Lorenz",
    title = "{Symmetries, safety, and self-supervision}",
    eprint = "2108.04253",
    archivePrefix = "arXiv",
    primaryClass = "hep-ph",
    doi = "10.21468/SciPostPhys.12.6.188",
    journal = "SciPost Phys.",
    volume = "12",
    number = "6",
    pages = "188",
    year = "2022"
}

@article{Bellagente:2020piv,
    author = {Bellagente, Marco and Butter, Anja and Kasieczka, Gregor and Plehn, Tilman and Rousselot, Armand and Winterhalder, Ramon and Ardizzone, Lynton and K{\"o}the, Ullrich},
    title = "{Invertible Networks or Partons to Detector and Back Again}",
    eprint = "2006.06685",
    archivePrefix = "arXiv",
    primaryClass = "hep-ph",
    doi = "10.21468/SciPostPhys.9.5.074",
    journal = "SciPost Phys.",
    volume = "9",
    pages = "074",
    year = "2020"
}

@article{Bellagente:2019uyp,
    author = "Bellagente, Marco and Butter, Anja and Kasieczka, Gregor and Plehn, Tilman and Winterhalder, Ramon",
    title = "{How to GAN away Detector Effects}",
    eprint = "1912.00477",
    archivePrefix = "arXiv",
    primaryClass = "hep-ph",
    doi = "10.21468/SciPostPhys.8.4.070",
    journal = "SciPost Phys.",
    volume = "8",
    number = "4",
    pages = "070",
    year = "2020"
}

@article{Kasieczka:2019dbj,
    author = "Butter, Anja and others",
    editor = "Kasieczka, Gregor and Plehn, Tilman",
    title = "{The Machine Learning landscape of top taggers}",
    eprint = "1902.09914",
    archivePrefix = "arXiv",
    primaryClass = "hep-ph",
    doi = "10.21468/SciPostPhys.7.1.014",
    journal = "SciPost Phys.",
    volume = "7",
    pages = "014",
    year = "2019"
}

@article{Heimel:2018mkt,
    author = "Heimel, Theo and Kasieczka, Gregor and Plehn, Tilman and Thompson, Jennifer M.",
    title = "{QCD or What?}",
    eprint = "1808.08979",
    archivePrefix = "arXiv",
    primaryClass = "hep-ph",
    doi = "10.21468/SciPostPhys.6.3.030",
    journal = "SciPost Phys.",
    volume = "6",
    number = "3",
    pages = "030",
    year = "2019"
}

\end{document}